\documentclass[journal,onecolumn,12pt,draftclsnofoot]{IEEEtran}
\usepackage{amssymb}
\usepackage{color,graphicx}
\usepackage{cite}
\usepackage{algorithm2e}
\usepackage[cmex10]{amsmath}
\usepackage{url}
\usepackage{amsopn}
\usepackage{authblk}

\newtheorem{definition}{Definition}

\newtheorem{corollary}{Corollary}
\newtheorem{proposition}{Proposition}

\newtheorem{example}{Example}
\newtheorem{remark}{Remark}
\newcommand{\remarkend}{ \IEEEQEDopen}

\allowdisplaybreaks

\begin{document}
\title{Message and State Cooperation in a Relay Channel When Only the Relay Knows the State\footnote[1] {The work of M. Li and A. Yener was supported in part by
the National Science
 Foundation under Grants CNS 0721445, CNS 0964364, CCF 0964362 and DARPA ITMANET Program under Grant
W911NF-07-1-0028. The work of O. Simeone was supported in part by
the National Science Foundation under Grant CCF 0914899. }
\footnote[2]{This work will be presented in part in Information
Theory and Applications Workshop, 2011.}}

\author{{Min Li$^1$, Osvaldo Simeone$^2$ and Aylin Yener$^1$}\\
\small {$^1$Dept. of Electrical Engineering, The Pennsylvania State
University, University Park, PA 16802}\\
\small {$^2$Dept. of Electrical
and Computer Engineering, New Jersey Institute of Technology, University Heights, NJ 07102}\\
\normalsize \textit{mxl971@psu.edu, osvaldo.simeone@njit.edu,
yener@ee.psu.edu}}

 \maketitle \vspace{-0.95in}
\begin{center}
February 1, 2011
\end{center}
\vspace{0.1in}

\begin{abstract}
\par A state-dependent relay channel is studied in which strictly causal channel state information
is available at the relay and no state information is available at
the source and destination. The source and the relay are connected
via two unidirectional out-of-band orthogonal links of finite
capacity, and a state-dependent memoryless channel connects the
source and the relay, on one side, and the destination, on the
other. Via the orthogonal links, the source can convey information
about the message to be delivered to the destination to the relay
while the relay can forward state information to the source. This
exchange enables cooperation between the source and the relay on
transmission of message and state information to the destination.
First, two
 achievable schemes are proposed that exploit both message and state cooperation. It is shown that a transmission scheme
 inspired by noisy network coding performs better than a strategy based on block Markov coding
 and backward decoding. Next, based on the given achievable schemes and appropriate upper bounds, capacity results
 are identified for some special cases. Finally, a Gaussian model is studied, along with corresponding numerical
 results that illuminate the relative merits of state and message
 cooperation.
\end{abstract}
\newpage

\section{Introduction}
\label{sec:introduction}
\par In a wireless network, main challenges to provide reliable communications include fading and interference. To
establish the fundamental performance limits of such channels, a
useful model is that the underlying channel is affected at each time
instant by a state variable, which is controlled by a certain state
distribution
\cite{shannon1958channels,gel1980coding,keshet2007channel}.
State-dependent channels are usually classified on the basis of the
availability of channel state information  at encoders and decoders.
Specifically, transmitting nodes may have no state information, or
else be informed about the state sequence in a strictly causal,
causal, or non-causal way \cite{keshet2007channel,LS_IZS2010,
LS_ISIT2010}. Causality refers to whether the state sequence at a
given time is known up to the previous instant (strictly causal
state information), up to and including the current time instant
(causal state information) or past, current and the future
(non-causal state information). For decoders, it is enough to
distinguish between the availability of state information or
not\cite{keshet2007channel}.

\par While fading induced state variations are often measured at the receivers
via training signals, leading to availability of the states at the
destinations, interference induced state variations are not so. In
particular, when the channel state models interference from other
users, the state can be more effectively measured at nodes that are
in the vicinity of the interferers, while nodes further away cannot
directly measure the state. In this case, it may happen that nodes
that are currently serving as transmitters may acquire state
information, while the respective receivers may not. In this paper,
we shall focus on such a scenario and analyze the performance
trade-off arising from the need to convey both message and state
information from transmitters to the receivers.

\par In previous work, capacity-achieving strategies have been proposed for
point-to-point memoryless channels with non-causal state information
\cite{gel1980coding, DPC_1983} and causal state information
\cite{shannon1958channels} at the encoder and no state information
at the decoder. These results, and the ones discussed throughout the
paper, assume that the state sequence is independently and
identically distributed (i.i.d.). Several multi-user channels have
also been widely investigated in similar settings including multiple
access channels (MACs)
\cite{sigurjonsson2005multiple,jafar2006capacity,philosof2007lattice,
somekh2008cooperative,kotagiri2008multiaccess,Permuter2010} and
relay channels\cite{mirmohseni2009compress, akhbari2010state,
zaidi2010cooperative, zaidi2010bounds}. References
\cite{sigurjonsson2005multiple,jafar2006capacity,philosof2007lattice,
somekh2008cooperative,kotagiri2008multiaccess} consider the MAC with
different availability of non-causal or causal state information at
some encoders. In \cite{Permuter2010}, message and state cooperation
is considered for a MAC with conferencing links with non-causal
state information available at the encoders and the decoder. For the
relay channel, reference \cite{zaidi2010cooperative} investigates
the case of non-causal state information at the relay, and proposes
a coding scheme that combines the strategies of decode-and-forward
\cite{cover1979capacity} and precoding against the state, while
reference \cite{mirmohseni2009compress} studies the case of causal
state information at the relay, and derives achievable rates by
combining the ideas of compress-and-forward \cite{cover1979capacity}
and adapting input codewords to the state (also known as Shannon
strategies \cite{shannon1958channels}).

\par This work also focuses on a state-dependent relay channel, but unlike previous work,
assumes that state information is available {\it only} at the relay
in a {\it strictly causal} fashion. This scenario is more relevant
in practical scenarios since in practice the state can be learned
only in a strictly causal way. For instance, in the case of an
interference network, an interfering sequence can be learned as it
is observed, and, thus, in a strictly causal manner. With strictly
causal state information, the strategies leveraged in
\cite{zaidi2010cooperative,mirmohseni2009compress}, for example, of
precoding against the state or Shannon strategies cannot be applied.
More fundamentally, the question arises as to whether strictly
causal, and thus outdated, state information may be useful at all in
a memoryless channel with i.i.d. state sequence. In fact, it is well
known that strictly causal state information is useless in
point-to-point channels. This conclusion can be seen along the lines
of argument for the fact that feedback does not increase the
capacity for memoryless point-to-point channels in
\cite{shannon1956zero}.

\par Recently, in \cite{LS_IZS2010, LS_ISIT2010}, it was found that, in contrast to the
case for point-to-point channels, for two-user MACs with independent
or common state information available strictly causally at the
encoders, capacity gains can be accrued by leveraging this
information. Our recent work \cite{lisimeoneyener2010} further
extended such results to MACs with arbitrary number of users by
proposing a coding scheme inspired by noisy network coding
\cite{lim2010noisy}. In \cite{LS_IZS2010,
LS_ISIT2010,lisimeoneyener2010 }, the main idea is to let each
transmitter convey a compressed version of the outdated state
information to the decoder, which in turn exploits this information
to perform partially coherent decoding. The results show that the
capacity region can be enlarged by allocating part of the
transmission resources to the transmission of the compressed state.

\par In this work, we consider a three-node relay channel where
the source and relay are connected via two out-of-band orthogonal
links of finite capacity, and a state-dependent memoryless channel
connects the source and relay, on one side, and the destination, on
the other. The source and destination have no state information,
while the relay has access to the state information in a strictly
causal manner. The channel model is shown in
Fig.~\ref{fig:model:relay}. This model is related to the class of
relay channels, that are not state-dependent, with orthogonal links
from the source to the relay and from the source and relay to the
destination investigated by El Gamal and Zahedi
\cite{elgamal2005capacity}. In fact, in the scenario under study, we
simplify the link from the source to the relay by modeling it as a
noiseless finite-capacity link, while adding a similar backward
relay-to-source link. Cooperation as enabled by orthogonal noiseless
links, also referred to as conferencing, was first introduced by
Willems \cite{willems1983discrete} for a two-user MAC channel and
was later extended to several settings \cite{dabora2006broadcast,
bross2008gaussian ,simeone2008three}. It is noted that, in practice,
orthogonal links can be realized if nodes are connected via a number
of different radio interfaces or wired links
\cite{gesbertmulti2010}.

\par
\begin{figure}[t]
\centering
\includegraphics[width=0.5\textwidth]{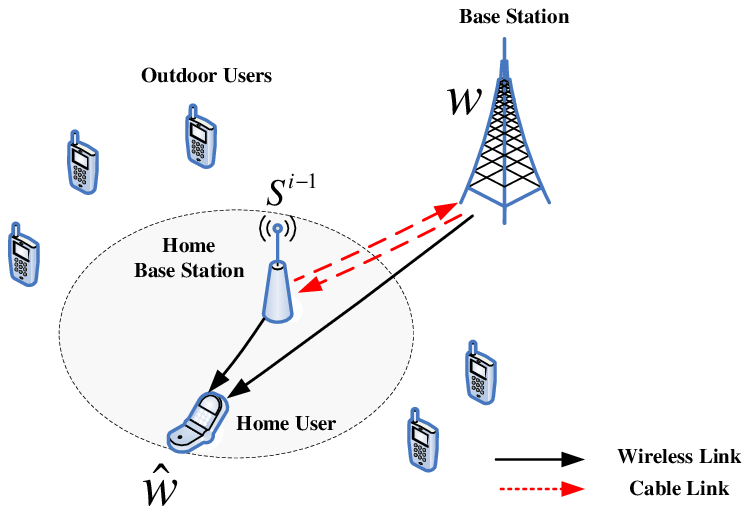}
\caption{The downlink transmission to a home user in a femtocell
provides an example application of the considered model illustrated
in Fig.~\ref{fig:model:relay}. The home base station is assumed to
be able to measure the interference from outdoor users.
}\label{fig:model:femtocell}
\end{figure}

 As an example, our model fits a downlink communication scenario
in a cellular network where femtocells are overlaid on a microcell
as shown in Fig.~\ref{fig:model:femtocell}. Femtocells are served by
home base stations, which are typically located around high
user-density hot spots, that can serve as intermediate nodes or
relays between users and the mobile operator network, to provide
better indoor voice service or data delivery for stationary or
low-mobility home users \cite{chandrasekhar2008femtocell}. The home
base station is typically connected to the outdoor base station via
an out-of-band wired link, e.g., a last-mile connection followed by
the Internet. The home base station may be able to measure the
interference created by outdoor users, whereas this may not be
possible at the base station or at indoor users. This gives rise to
the system model we consider in this paper, as can be readily
observed from Figs.~\ref{fig:model:femtocell} and
\ref{fig:model:relay}.

\par In the considered model, cooperation between source and relay
through the conferencing links can aim at two distinct goals: $i$)
Message transmission: Through the source-to-relay link, the source
can provide the relay with some information about the message to be
conveyed to the destination, thus enabling message cooperation;
$ii$) State transmission: Through the relay-to-source link, the
relay can provide the source with some information about the state,
thus enabling cooperative transmission of the state information to
the destination. We propose two achievable schemes, one based on
conventional block Markov coding \cite{cover1979capacity} and
backward decoding \cite{willems1982informationtheoretical} and one
inspired by noisy network coding. We show that the latter
outperforms the former in general. Moreover, based on these
achievable rates, we identify capacity results for some special
cases of the considered model. We also investigate the optimal
capacity allocation between the source-to-relay and relay-to-source
links where the total conferencing capacity is fixed. Finally, we
derive achievable rates and some capacity results for the Gaussian
version of the system at hand and elaborate on numerical results.

\par The remainder of this paper is organized as follows. Section
\ref{sec:channel:model:relay} formally describes the relay model
considered in this work. Section \ref{sec:relay:DM:achievable}
illustrates two different achievable coding schemes and presents the
resulting achievable rates. Section \ref{sec:relay:DM:capacity}
identifies capacity results for some special cases. Section
\ref{sec:relay:resource:fixed} studies the scenario in which the
total conferencing capacity is fixed and elaborates on optimal
capacity allocation. Section \ref{sec:relay:Gaussian:numerical}
studies the Gaussian case of our model and provides numerical
results. Section \ref{sec:relay:conclusion} concludes the work.

\par Throughout the paper the following notation is used. Probability distributions are denoted by $p$ subscripted by the random variables involved,
e.g., $p_X \left( x \right)$ is the probability of $ X = x$, $
p_{Y\left| X \right.} \left( {y\left| x \right.} \right)$ is the
conditional probability of $ Y = y $ given $X = x$, etc. We will
drop subscripts from the probability functions when the meaning is
clear from the context, e.g., $p\left( x \right)$ stands for $p_X
\left( x \right)$. Also ${ x}^i$ denotes vector $\left[ {x_{1}
,...,x_{i} } \right]$. $\mathbb{E}\left[ X \right]$ denotes the
expectation of random variable $X$. ${\cal N}\left( {0,\sigma ^2 }
\right)$ denotes a zero-mean Gaussian distribution with variance
$\sigma^2$. ${\cal C}(x)$ is defined as ${\cal C}\left( x \right) =
\frac{1}{2}\log _2 \left( {1 + x} \right)$.

\section{System Model} \label{sec:channel:model:relay}
\par
\begin{figure}[t]
\centering
\includegraphics[width=0.75\textwidth]{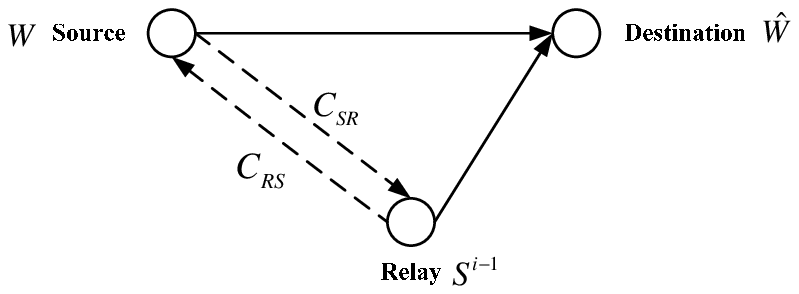}
\caption{A state-dependent relay channel with two unidirectional
out-of-band orthogonal links.}\label{fig:model:relay}
\end{figure}

 In this section, we present the channel model and provide relevant definitions.
As depicted in Fig.~\ref{fig:model:relay}, we study a three-node
relay channel where the source and relay are connected via two
unidirectional out-of-band orthogonal links of finite capacity,
while there is a state-dependent memoryless channel between the
source and relay, on one side, and the destination, on the other.
Note that the relay transmits and receives simultaneously over two
orthogonal channels.

\par The channel is characterized by the tuple:
\begin{align}
\left( {{\cal X}  \times {\cal X}_R ,{\cal S},{\cal Y},p\left( s
\right),p\left( {y\left| {s, x ,x_R} \right.} \right)}, C_{SR},
C_{RS}\right)
\end{align}
with source input alphabet ${\cal X}$, relay input alphabet ${\cal
X}_R $, destination output alphabet ${\cal Y}$ and channel state
alphabet ${\cal S}$. The capacity per channel use of the
source-to-relay and relay-to-source out-of-band, also known as
conferencing \cite{willems1983discrete}, links are given by
$C_{SR}$, $C_{RS}$ respectively. The state sequence is assumed to be
i.i.d., i.e., $ p\left( {s^n } \right) = \prod\limits_{i = 1}^n
{p\left( {s_i } \right)} $. The relay channel is discrete memoryless
(DM) in the sense that at any discrete time $i = 1,...,n$, we have
\begin{align}
p\left( {y_i \left| {s^i, x^i ,x_R^i,y^{i - 1} } \right.} \right) =
p\left( {y_i \left| {s_i, x_i ,x_{R,i} } \right.} \right).
\end{align}
We assume that state information is available to the relay in a {\it
strictly causal} manner while there is no state information at the
source and destination.
\begin{definition}
\label{def:randomcode} Let $W$, uniformly distributed over the set
${\cal W} =[1:2^{nR}]$, be the message sent by the source. A
$(2^{nR}, n)$ code consists of:
\begin{enumerate}
  \item Conferencing codes: Conferencing mappings are defined as
  \begin{align}
&h_{SR,i} : {\cal W} \times {\cal T}_{RS}^{i - 1}  \to {\cal T}_{SR,i}, \label{equ:conf:code:SR}\\
&h_{RS,i} : {\cal S}^{i - 1} \times {\cal T}_{SR}^{i - 1}   \to
{\cal T}_{RS,i}, \label{equ:conf:code:RS}
\end{align}
where \eqref{equ:conf:code:SR} generates the $i$th symbol sent on
the source-to-relay link based on the message and all symbols
previously received on the relay-to-source link, while
\eqref{equ:conf:code:RS} generates the $i$th symbol sent on the
relay-to-source link based on the strictly causal states and all
symbols previously received on the source-to-relay link. Note that
${\cal T}_{SR,i}$ is the alphabet of the conferencing message sent
from the source to the relay, while ${\cal T}_{RS,i}$ is the
alphabet of the conferencing message sent from the relay to the
source at time instant $i$, $i = 1,...,n$. Such mappings are
permissible if the following capacity-conserving conditions are
satisfied:
\begin{align}
&\frac{1} {n}\sum\limits_{i = 1}^{n } {\log _2 \left| {{\cal
T}_{SR,i} }
\right|}  \le C_{SR}, \label{equ:conf:code:CSR}\\
&\frac{1}{n}\sum\limits_{i = 1}^{n } {\log _2 \left| {{\cal
T}_{RS,i} } \right|}  \le C_{RS}. \label{equ:conf:code:CRS}
\end{align}

  \item Encoder mappings at the source:
  \begin{align}
f_i : {\cal W} \times {\cal T}_{RS}^{i}  \to {\cal X}_i ,\forall\; i
= 1,...,n,
\end{align}
which generates the channel input at the source at time $i$ based on
the message and the information received from the relay up to and
including time $i$ on the relay-to-source link.
  \item Encoder mappings at the relay:
  \begin{align}
  f_{R,i} :  {\cal S}^{i - 1} \times {\cal T}_{SR}^{i}  \to {\cal
  X}_{R,i},\forall\; i = 1,...,n,
  \end{align}
  which generates the channel input at the relay at time $i$ based
  on the strictly causal state information and the information
  received from the source up to and including time $i$ on the source-to-relay link.
  \item Decoder mapping at the destination:
  \begin{align}
  g: {\cal Y}^n  \to {\cal W},
\end{align}
which produces the estimate of message at the destination based on
the received sequences.
\end{enumerate}
\end{definition}

\par The average probability of error, $\Pr(E)$, is defined by:
\begin{align} \label{equ:prob:error}
\Pr (E) = \frac{1} {{2^{nR} }}\sum\limits_{w = 1}^{2^{nR} } {\Pr
\left( {g\left( {y^n } \right) \ne w \left| { w \:\:sent} \right.}
\right)}.
\end{align}
\par A rate $R$ is achievable if there exists a
sequence of codes $(2^{nR}, n)$ as defined above such that the
probability of error $\Pr(E) \to 0$ as $n \to \infty$. The capacity
of this channel is the supremum of the set of all achievable rates.

\section{Achievable Schemes and Upper Bound}
\label{sec:relay:DM:achievable}
\par In this section, we demonstrate two different coding schemes
that exploit the potential benefits of message and state cooperation
between source and relay. We also identify a simple upper bound on
the capacity.

\subsection{Scheme 1: Block-based Message and State Cooperation} We
first propose an achievable scheme based on conventional block
Markov coding and backward decoding.
\begin{proposition}\label{proposition:achievable1}
\par For the DM state-dependent relay channel of
Fig.~\ref{fig:model:relay}, any non-negative rate smaller than $R_1$
is achievable where
\begin{align}
R_1  = \mathop {\max }\limits_{{\cal P}_1 } \min \left(
\begin{array}{l}
 I\left( {X;Y\left| {X_R ,V,U} \right.} \right) + C_{SR} , \\
 I\left( {X,X_R ;Y\left| V \right.} \right) - I\left( {V;S\left| Y \right.} \right), \\
 I\left( {X,X_R ;Y\left| {V,U} \right.} \right) + C_{SR}  + C_{RS}  - I\left( {V;S\left| Y \right.}
 \right)
 \end{array} \right)\label{equ:achievable:1}
\end{align}
with the maximum taken over the distributions in the set of
\begin{align}
{\cal P}_1  = \left\{ {p\left( {v,u,s,x,x_R,y} \right):p\left( s
\right)p\left( {v\left| s \right.} \right)p\left( u \right)p\left(
{x\left| u \right.} \right)p\left( {x_R \left| u \right.}
\right)p\left( {y\left| {s,x,x_R } \right.} \right)} \right\}
\label{equ:inputd:1}
\end{align}
subject to the constraint:
\begin{align}
 I\left( {X_R; Y\left| X, U \right.} \right) + \min \left( C_{RS}, I\left( {X, U ;Y} \right) \right)  \ge I\left( {V;S\left| Y \right.}
\right).
  \label{equ:constraint1:th1}
 \end{align}
\end{proposition}
\begin{IEEEproof}[Sketch of Proof]
\label{proof:proposition:achievablerate1} The idea is to follow a
natural block Markov strategy. Specifically, the message $w$ is
split by the source into $\left(b-1\right)$ parts, $\left( {w_1
,...,w_{b - 1} } \right)$, $w_j  \in \left[ {1:2^{nR_1 } } \right]$,
$j=1,...,\left(b-1\right)$, which are transmitted over $b$ blocks,
each block consisting of $n$ channel uses where $n = \frac{m}{b}$,
and $m$ is the number of total channel uses. At the end of each
block, the relay compresses the state sequence that has affected the
channel over the block with the aim of conveying such information to
the destination in the next block. Compression exploits the side
information at the destination via Wyner-Ziv
coding\cite{wynerziv1976}. Conferencing takes place before the
beginning of each block. Specifically, through conferencing, before
the $j$th block, the source conveys part of the message $w_j$ to the
relay in order to enable message cooperation, while the relay sends
part of the bin index produced by Wyner-Ziv coding
\cite{wynerziv1976, cover1979capacity} to the source to enable
cooperative state transmission. The exchange state and message
information is sent cooperatively by the source and relay, while the
remaining part of the message $w_j$ is sent independently by the
source and the remaining part of the bin index is sent by the relay
alone. This strategy is referred to as block-based message and state
cooperation. Decoding takes place by backward decoding
\cite{willems1982informationtheoretical}. Specifically, starting
from the last reception, the destination first retrieves the
compressed state information for block $\left(b-1\right)$. After
that, it performs coherent decoding to recover message $w_{b-1}$
from the $\left(b-1\right)$th block reception by exploiting the
state information retrieved. Using the decoded message, the
destination turns to retrieve the compressed state information for
block $\left(b-2\right)$, and then decodes the corresponding message
$w_{b-2}$. Repeating this operation until back to the first block,
the destination recovers all the messages over blocks. Details of
the proof are provided in Appendix
\ref{appendix:proposition:achievable1}.
\end{IEEEproof}

\begin{remark} \label{remark:achievable1}
\par To interpret \eqref{equ:achievable:1} to \eqref{equ:constraint1:th1} in light of the transmission strategy discussed above for scheme 1, we
remark that $V$ represents the compressed state information and $U$
accounts for the codeword transmitted cooperatively by the source
and relay, which conveys both state and message information they
share. Bound \eqref{equ:constraint1:th1} imposes that the Wyner-Ziv
rate $I\left( {V;S\left| Y \right.} \right)$ is supported by the
cooperative transmission of the source and relay, whose rate is
limited by $\min \left( {C_{RS} ,I\left( {X,U;Y} \right)} \right)$
and the information sent independently from the relay $I\left( {X_R
;Y\left| {X,U} \right.} \right)$. The mutual information terms in
\eqref{equ:achievable:1}, and in particular the conditioning on $V$,
account for the fact that the destination has information about the
channel via the compressed state $V$, which allows for partial or
complete coherent decoding. Moreover, the second and third term in
\eqref{equ:achievable:1} reflect the cost in terms of rate to be
paid for the transmission of compressed state information.\remarkend
\end{remark}

\subsection{Scheme 2: Burst Message Cooperation and Block-based State
Cooperation} In this subsection, we propose a second transmission
scheme inspired by noisy network coding \cite{lim2010noisy}.

\begin{proposition}\label{proposition:achievable2}
\par For the DM state-dependent relay channel of
Fig.~\ref{fig:model:relay}, any non-negative rate smaller than $R_2$
is achievable where
\begin{align}  \label{equ:achievable:2}
R_2  = \mathop {\max }\limits_{{\cal P}_2 } \min \left(
\begin{array}{l}
 I\left( {X;Y\left| {X_R ,V,U} \right.} \right) + C_{SR} , \\
 I\left( {X,X_R,V,U;Y} \right) - I\left( {V;S\left| X_R, U \right.} \right), \\
 I\left( {X,X_R,V ;Y\left| {U} \right.} \right) + C_{SR}  + C_{RS}  - I\left( {V;S\left| X_R, U \right.}
 \right)
 \end{array} \right)
\end{align}
with the maximum taken over the distributions in the set of
\begin{align}
{\cal P}_2  = \left\{ {p\left( {v,u,s,x,x_R,y} \right):p\left( s
\right)p\left( {v\left| {s,x_R ,u} \right.} \right)p\left( u
\right)p\left( {x\left| u \right.} \right)p\left( {x_R \left| u
\right.} \right)p\left( {y\left| {s,x,x_R } \right.} \right)}
\right\}. \label{equ:inputd:2}
\end{align}
\end{proposition}

\begin{IEEEproof}[Sketch of Proof]
\label{proof:proposition:achievablerate2} Inspired by the noisy
network coding scheme in \cite{lim2010noisy}, the same message $w$,
$w \in \left[ {1:2^{nbR_2 } } \right]$, is sent at the source over
all $b$ blocks of transmission with each consisting of $n$ channel
uses. Thus, unlike scheme 1 discussed above, here information
exchange about the message between source and relay takes place only
one at the beginning of the first block. This way, the source shares
part of the message $w$ with the relay in order to enable message
cooperation. As for the state, at the end of each block, the relay
compresses the state sequence over the block {\it without explicit
Wyner-Ziv coding}, that is, without binning \cite{lim2010noisy}.
Exchange of state information between relay and source takes place
before the beginning of each block as for scheme 1 proposed above.
Source and relay cooperatively send the message and state
information they share, while the source sends the remaining part of
the message independently and the relay sends the remaining part of
the compression index alone for each block. This transmission scheme
is referred to as burst message cooperation and block-based state
cooperation strategy. At the end of $b$ blocks of transmission, the
destination performs {\it joint decoding over all blocks} of
reception {\it without explicitly decoding the compressed state
information} as for the noisy network coding scheme
\cite{lim2010noisy}. Details of the proof are provided in Appendix
\ref{appendix:proposition:achievable2}.
\end{IEEEproof}

\begin{remark} \label{remark:achievable2}
\par To interpret \eqref{equ:achievable:2} to \eqref{equ:inputd:2} in light of the transmission strategy discussed above
 and in comparison the one in scheme 1, we point out that, as in
Remark \ref{remark:achievable1}, $V$ represents the compressed state
information while $U$ denotes for the common message and state
information. Each mutual information term in
\eqref{equ:achievable:2}, in particular the conditioning on $V$, has
for a similar interpretation as explained in Remark
\ref{remark:achievable1}. Unlike scheme 1, however, the compressed
state $V$ is generated without explicit Wyner-Ziv coding and without
requiring correct decoding of the compressed state at the receiver.
This fact, as detailed in the proof, makes it possible to choose $V$
to be dependent of $X_R$, $U$ and $S$, instead of only $S$ in scheme
1. Moreover, the rate loss due to the need to convey state
information can be smaller than $I\left(V;S\left|Y\right.\right)$ in
\eqref{equ:achievable:1}, as discussed in Proposition
\ref{proposition:scheme2:beat:scheme1}. Finally, since the decoding
is implemented jointly without recovering all the compressed states
correctly in scheme 2, there is no explicit additional constraint
\eqref{equ:constraint1:th1}.\remarkend
\end{remark}
\subsection{Comparison of Achievable Rates} Based on the discussion
above, we expect scheme 2 of Proposition
\ref{proposition:achievable2} to outperform scheme 1 of Proposition
\ref{proposition:achievable1}. This is shown by the following
proposition.
\begin{proposition} \label{proposition:scheme2:beat:scheme1}
$R_2  \ge R_1$.
\end{proposition}
\begin{IEEEproof} We prove the results by showing that the three terms in \eqref{equ:achievable:2} are larger or
equal than the ones in \eqref{equ:achievable:1}. This, coupled with
the fact that the characterization of $R_2$ does not have additional
constraint \eqref{equ:constraint1:th1} and with the more general
distribution $p\left(v\left| s, x_R,u \right.\right)$ allowed by
scheme 2 over scheme 1 (which constrains the distribution as
$p\left(v\left|s\right.\right)$), is enough to conclude the proof.
Specifically, setting $p\left(v\left| s, x_R,u \right.\right) =
p\left({v\left| s\right.}\right)$ in ${\cal P}_2$, we have that:
\begin{enumerate}
  \item The first term in \eqref{equ:achievable:2} is the same as the first term in
  \eqref{equ:achievable:1}.
  \item The second terms are also equal since
\begin{align}
  &I\left( {X,X_R ,V,U;Y} \right) - I\left( {V;S\left| {X_R, U } \right.} \right)\\
  &= I\left( {X,X_R;Y\left| V \right.} \right) + I\left(V;Y\right)- I\left( {V;S\left| {X_R, U } \right.}
  \right) \label{equ:com:markov:chain:new}\\
  &= I\left( {X,X_R;Y\left| V \right.} \right) + H\left( {V\left| S \right.} \right) - H\left( {V\left| Y \right.} \right) \label{equ:com:independent}\\
  &= I\left( {X,X_R;Y\left| V \right.} \right) - I\left( {V;S\left| Y \right.}
  \right), \label{equ:com:markov:chain}
 \end{align}
 where \eqref{equ:com:markov:chain:new} follows from the Markov chain $
 U \leftrightarrow \left(X,X_R,V\right) \leftrightarrow Y$ for the distribution considered, \eqref{equ:com:independent} is because $V$ is independent of $\left(U, X_R\right)$, and \eqref{equ:com:markov:chain} follows from the Markov chain $
V \leftrightarrow S \leftrightarrow Y$.
  \item The third term of \eqref{equ:achievable:2} is larger or equal than the corresponding term in \eqref{equ:achievable:1} since
  \begin{align}
  &I\left( {X,X_R ,V;Y\left| U \right.} \right)+ C_{SR}  + C_{RS} - I\left( {V;S\left| {X_R, U } \right.} \right) \\
  &= I\left( {X,X_R ;Y\left| {V,U} \right.} \right) + C_{SR}  + C_{RS}  + I\left( {V;Y\left| U \right.} \right) - I\left( {V;S\left| {X_R, U } \right.} \right) \\
  &= I\left( {X,X_R ;Y\left| {V,U} \right.} \right) + C_{SR}  + C_{RS}  + H\left( {V\left| S \right.} \right) - H\left( {V\left| {Y,U} \right.} \right) \label{equ:com:independent2} \\
  &\ge I\left( {X,X_R ;Y\left| {V,U} \right.} \right) + C_{SR}  + C_{RS}  + H\left( {V\left| S \right.} \right) - H\left( {V\left| Y \right.} \right) \label{equ:com:conditioning} \\
  &= I\left( {X,X_R ;Y\left| {V,U} \right.} \right) + C_{SR}  + C_{RS}  - I\left( {V;S\left| Y \right.}
  \right), \label{equ:com:markov:chain2}
  \end{align}
  where \eqref{equ:com:independent2} is because $V$ is independent of $\left(U, X_R\right)$, \eqref{equ:com:conditioning} holds because conditioning reduces entropy, while \eqref{equ:com:markov:chain2} again follows from the Markov chain $
V \leftrightarrow S \leftrightarrow Y$.
\end{enumerate}
\end{IEEEproof}
\subsection{An Upper Bound} Here we derive a simple upper bound.
\begin{proposition}
\par For the DM state-dependent relay channel of
Fig.~\ref{fig:model:relay}, the capacity is upper bounded by
\begin{align}  \label{equ:simple:upperbound}
R_{upp}  = \mathop {\max }\limits_{{\cal P}_{upp} } \min \left(
{I\left( {X,X_R ;Y} \right),I\left( {X;Y\left| {X_R ,S} \right.}
\right) + C_{SR} } \right)
\end{align}
with the maximum taken over the distributions in the set of
\begin{align}
{\cal P}_{upp}  = \left\{ {p\left( {s,x,x_R,y} \right):p\left( s
\right)p\left( {x,x_R } \right)p\left( {y\left| {s,x,x_R } \right.}
\right)} \right\}. \label{equ:inputd:out}
\end{align}
\end{proposition}
\begin{IEEEproof} The upper bound \eqref{equ:simple:upperbound} is essentially a cut-set bound \cite{info_cover}, where the
first term corresponds to the MAC cut between source-relay and
destination, and the second is the cut between source and
relay-destination. Given presence of the state sequence, calculation
requires some care and is detailed below.
\par For the first term, consider a
genie-aided system in which the message is also provided to the
relay and the state $s^{i-1}$ is also provided to the source at time
$i$. The system can be now seen as being point-to-point with inputs
$\left(X, X_R\right)$, output $Y$ and with strictly causal state
information. In this case, it is well known that state information
does not increase capacity, which is given by the first term in
\eqref{equ:simple:upperbound}. The result can also be seen from the
Fano' inequality \cite{info_cover} as
\begin{align}
 R_{upp} &\le \frac{1}{n}I\left( {W;Y^n } \right) +  \epsilon_n \\
  &\le \frac{1}{n}\sum\limits_{i = 1}^n {I\left( {W;Y_i \left| {Y^{i - 1} } \right.} \right)}  + \epsilon_n \\
  &\le \frac{1}{n}\sum\limits_{i = 1}^n {I\left( {W,Y^{i - 1} ,X_i ,X_{R,i} ;Y_i } \right)}  +  \epsilon_n \label{equ:upper:b1} \\
  &\le \frac{1}{n}\sum\limits_{i = 1}^n {I\left( {X_i ,X_{R,i} ;Y_i } \right)}
  + \epsilon_n \label{equ:upper:b2}
\end{align}
with $\epsilon_n \to 0$ as $n \to \infty$, where
\eqref{equ:upper:b1} follows from the non-negativity of mutual
information, \eqref{equ:upper:b2} follows from the Markov chain $
\left( {W,Y^{i - 1} } \right) \leftrightarrow \left( {X_i ,X_{R,i} }
\right) \leftrightarrow Y_i$. This Markov chain can be seen as a
consequence of the independence of $S_i$ and $\left(W,
Y^{i-1}\right)$, and the Markov chain $\left(W, Y^{i-1}\right)
\leftrightarrow \left(X_i, X_{R,i}, S_i\right) \leftrightarrow Y_i$.
\par For the second term, consider another genie-aided system in which the perfect state
information is provided to the destination. Then, by the Fano'
inequality \cite{info_cover}, we have
\begin{align}
 R_{upp}  &\le \frac{1}{n}I\left( {W;Y^n ,S^n ,T_{SR}^n } \right) + \epsilon_n \\
  &= \frac{1}{n}I\left( {W;Y^n ,T_{SR}^n \left| {S^n } \right.} \right) +  \epsilon_n \label{equ:upper:b3}\\
  &= \frac{1}{n}I\left( {W;Y^n \left| {S^n, T_{SR}^n  } \right.} \right) + \frac{1}{n}I\left( {W;T_{SR}^n \left| {S^n } \right.} \right)
  + \epsilon_n \label{equ:upper:b4}
\end{align}
with $\epsilon_n \to 0$ as $n \to \infty$, where
\eqref{equ:upper:b3} holds because $W$ and $S^n$ are independent and
\eqref{equ:upper:b4} follows from the chain rule. Note that
\begin{align}
 &\frac{1}{n}I\left( {W;T_{SR}^n \left| {S^n } \right.} \right) \nonumber \\
  &\le \frac{1}{n}H\left( {T_{SR}^n } \right) \\
  &\le \frac{1}{n}\sum\limits_{i = 1}^n {H\left( {T_{SR,i} } \right)}  \label{equ:upper:b5}\\
  &\le C_{SR}, \label{equ:upper:b6}
\end{align}
where \eqref{equ:upper:b5} is due to the fact that conditioning
reduces entropy and \eqref{equ:upper:b6} follows from the definition
of permissible conferencing mapping given by
\eqref{equ:conf:code:CSR}. Moreover, we have that
\begin{align}
  &\frac{1}{n}I\left( {W;Y^n \left| {S^n ,T_{SR}^n } \right.} \right) \nonumber \\
  &= \frac{1}{n}\sum\limits_{i = 1}^n {I\left( {W;Y_i \left| {Y^{i - 1} ,S^n ,T_{SR}^n } \right.} \right)}  \\
  &= \frac{1}{n}\sum\limits_{i = 1}^n {I\left( {W;Y_i \left| {Y^{i - 1} ,S^n } \right.,T_{SR}^n ,X_R^n } \right)}  \label{equ:upper:b7}\\
  &= \frac{1}{n}\sum\limits_{i = 1}^n {\left[ {H\left( {Y_i \left| {Y^{i - 1} ,S^n } \right.,T_{SR}^n ,X_R^n } \right) - H\left( {Y_i \left| {W,Y^{i - 1} ,S^n } \right.,T_{SR}^n ,X_R^n } \right)} \right]}  \\
  &= \frac{1}{n}\sum\limits_{i = 1}^n {\left[ {H\left( {Y_i \left| {Y^{i - 1} ,S^n } \right.,T_{SR}^n ,X_R^n } \right) - H\left( {Y_i \left| {W,Y^{i - 1} ,S^n } \right.,T_{SR}^n ,X_R^n ,T_{RS}^n ,X^n } \right)} \right]}  \label{equ:upper:b8} \\
  &\le \frac{1}{n}\sum\limits_{i = 1}^n {\left[ {H\left( {Y_i \left| {X_{R,i} ,S_i } \right.} \right) - H\left( {Y_i \left| {X_i ,X_{R,i} ,S_i } \right.} \right)} \right]}  \label{equ:upper:b9} \\
  &= \frac{1}{n}\sum\limits_{i = 1}^n {I\left( {X_i ;Y_i \left| {X_{R,i} ,S_i } \right.}
  \right)},
\end{align}
where \eqref{equ:upper:b7} holds because $X_{R,i}$ is a function of
$\left( {T_{SR}^i ,S^{i - 1} } \right)$, \eqref{equ:upper:b8} holds
because $T_{RS,i}$ is a function of $ \left( {T_{SR}^{i - 1} ,S^{i -
1} } \right)$ while $X_i$ is a function of $\left( {W,T_{RS}^{i} }
\right)$, \eqref{equ:upper:b9} follows from the memoryless property
of the channel and the fact that conditioning reduces entropy.
Overall, we have
\begin{align}
R_{upp}  \le \frac{1}{n}\sum\limits_{i = 1}^n {I\left( {X_i ;Y_i
\left| {X_{R,i} ,S_i } \right.} \right)}  + C_{SR}   +  \epsilon_n
\label{equ:upper:b10}
\end{align}
with $\epsilon_n \to 0$ as $n \to \infty$.

\par Finally, from \eqref{equ:upper:b2} and \eqref{equ:upper:b10},
the proof is concluded using the standard approach of introducing a
time-sharing variable $Q$ uniformly distributed in the set
$\left[1:n\right]$ and then arguing that one can set $Q$ to be
constant without loss of optimality {\cite[Ch.15]{info_cover}}.
\end{IEEEproof}

\section{Special Cases and Capacity Results} \label{sec:relay:DM:capacity}
 In this section, we consider three special cases of the general model studied above, namely: $i)$ No message and state
 cooperation, in which $C_{SR} = C_{RS} = 0$; $ii)$ Message cooperation only, in which $ C_{SR}> 0, C_{RS} = 0$;
 $iii)$ State cooperation only, in which $C_{SR} = 0, C_{RS}> 0$. We establish capacity results
 for a special class of channels for each case.

\subsection{No Message and State Cooperation}
\begin{corollary}
If $C_{SR} = C_{RS} = 0$, any non-negative rate smaller than
$R_{21}$ is achievable where
\begin{align} \label{equ:achievable:21}
R_{21}  = \mathop {\max }\limits_{{\cal P}_{21} } \min \left(
{I\left( {X;Y\left| {X_R ,V} \right.} \right),I\left( {X,X_R ,V;Y}
\right) - I\left( {V;S\left| {X_R } \right.} \right)} \right)
\end{align}
with the maximum taken over the distributions in the set of
\begin{align}
 {\cal P}_{21}  = \left\{ {p\left({v,s,x,x_R,y}\right) :p\left(s\right) p\left({v\left| s, x_R
\right.}\right) p\left({x}\right) p\left({x_R}\right)
p\left({y\left| {s,x,x_R } \right.}\right) } \right\}.
\label{equ:inputd:21}
\end{align}
\end{corollary}
\begin{IEEEproof} The achievable rate follows from $R_2$
\eqref{equ:achievable:2} by setting $C_{SR} = C_{RS} = 0$ and $U =
\emptyset$, since no information is shared between the source and
relay.
\end{IEEEproof}
\par This rate turns out to be optimal, i.e., capacity-achieving, for a special class
of relay channels, which includes modulo-additive state-dependent
relay channels, see Example \ref{example:det:relay}.

\begin{proposition} \label{proposition:det:relay}
\par Let ${\cal P}_{21}^*$ denote the set of distributions defined by:
\begin{align} \label{equ:distribution:nocoop}
 {\cal P}_{21}^*  = \left\{ {p\left({s,x,x_R,y}\right) : p\left(s\right) p\left({x}\right) p\left({x_R}\right) p\left({y\left| {s,x,x_R } \right.}\right) }
 \right\}.
\end{align}
If $C_{SR} = C_{RS} = 0$,
\begin{align}
&H\left( {Y\left| {X,X_R ,S} \right.} \right) = 0 , \label{det:relay:cond1}\\
{\rm and}\;&H\left( {S\left| {X,X_R, Y } \right.} \right) = 0
\label{det:relay:cond2}
\end{align}
are satisfied for all distributions in ${\cal P}_{21}^*$, then the
capacity is given by:
\begin{align}
C_{21}  = \mathop {\max }\limits_{{\cal P}_{21}^* } \min \left(
{H\left( {Y\left| {X_R ,S} \right.} \right),I\left( {X,X_R ;Y}
\right)} \right). \label{equ:capacity:C21}
\end{align}

\end{proposition}

\begin{IEEEproof} The achievability is straightforward by setting $V = S$ and applying
assumptions \eqref{det:relay:cond1} and \eqref{det:relay:cond2} when
evaluating \eqref{equ:achievable:21}. Specifically, we have
\begin{align}
 &I\left( {X;Y\left| {X_R ,S} \right.} \right)= H\left( {Y\left| {X_R ,S} \right.}
 \right), \label{equ:capacity:C21:achievable1}
 \end{align}
 and
 \begin{align}
 &I\left( {X,X_R ,S;Y} \right) - I\left( {S;S\left| {X_R } \right.} \right) \\
  &= I\left( {X,X_R ;Y} \right) + I\left( {S;Y\left| {X,X_R } \right.} \right) - H\left( S \right)\\
  &= I\left( {X,X_R ;Y} \right) - H\left( {S\left| {X,X_R, Y } \right.} \right)\\
  &= I\left( {X,X_R ;Y} \right). \label{equ:capacity:C21:achievable2}
\end{align}
\par To obtain a converse result, we follow from \eqref{equ:simple:upperbound} and
note the fact that $X$ and $X_R$ must be independent since source
and relay cannot cooperate when $C_{SR} = C_{RS} = 0$. Hence, the
capacity is upper bounded by \eqref{equ:simple:upperbound} evaluated
for some product input distribution
$p\left(x\right)p\left(x_R\right)$. Overall, we have:
\begin{align}
C_{21}  &\le I\left( {X;Y\left| {X_R ,S} \right.} \right) = H\left(
{Y\left| {X_R ,S} \right.} \right),
\label{equ:nomessage:state:coop:bound1}\\
C_{21}  &\le I\left( {X,X_R ;Y} \right)
\label{equ:nomessage:state:coop:bound2}
\end{align}
for some input distribution $p\left(x\right)p\left(x_R\right)$. The
proof is concluded by maximizing the mutual information terms
\eqref{equ:nomessage:state:coop:bound1} and
\eqref{equ:nomessage:state:coop:bound2} over the same input
distribution $p\left(x\right)p\left(x_R\right)$.
\end{IEEEproof}

\begin{remark}
Achievability of the capacity
\eqref{equ:distribution:nocoop}$-$\eqref{equ:capacity:C21} has been
proved above via scheme 2. The same capacity result {\it cannot} be
obtained by setting $U = \emptyset, V = S$ in $R_1$ from scheme 1 of
Proposition \ref{proposition:achievable1}, since we have the
additional constraint $I\left( {X_R ;Y\left| X \right.} \right) \ge
H\left(S \left|Y\right.\right) $. This points to the advantage of
noisy network coding-like strategy used by scheme 2. \remarkend
\end{remark}

\begin{remark}\label{remark:det:interpretation}
\par Condition \eqref{det:relay:cond1}
basically states that, when fixed $X$ and $X_R$, there is no other
source of uncertainty in the observation $Y$ beside the state $S$.
Condition \eqref{det:relay:cond2}, instead, says that the state $S$
is perfectly determined when $Y, X$ and $X_R$ are known. These
conditions guarantee that providing information about the state
directly reduces the uncertainty about the input $X$ and $X_R$. The
fact that the relay can increase the achievable rate up to
$I\left(X, X_R; Y\right)$ in \eqref{equ:capacity:C21} can be
interpreted in light of this fact since the relay signal $X_R$
directly contributes to the achievable rate even though the relay is
not aware of the message. This will be further discussed in Remark
\ref{remark:state:feedback:coding} for a Gaussian model. \remarkend
\end{remark}

\begin{example} \label{example:det:relay}
\par Consider a binary modulo-additive state-dependent relay channel defined by
\begin{align}
Y = X  \oplus X_R  \oplus S,
\end{align}
where $S \sim Bernoulli\left( p_s \right)$. Let us further impose
the cost constraints on the source and relay codewords $\left(x^n,
x_{R}^n\right)$,
\begin{align}
\frac{1}{n}\sum\limits_{i = 1}^n {{\mathbb E} \left[ X_{i}\right] }
\le p, \frac{1}{n}\sum\limits_{i = 1}^n { {\mathbb E} \left[
X_{R,i}\right] } \le p_r
\end{align}
with $0 \le p, p_r \le \frac{1}{2}$. Extending the capacity result
of Proposition \ref{proposition:det:relay} to channels with cost
constraints is straightforward and leads simply to limiting the set
of feasible distributions \eqref{equ:distribution:nocoop} by
imposing the constraints that ${{\mathbb E} \left[ X\right] } \le p$
and ${{\mathbb E} \left[ X_R\right] } \le p_r$, see, e.g.,
{\cite{YKGamalLecture}}. Therefore the capacity is given by:
\begin{align}
C_{\rm bin} = \min\left(H_b(p), H_b \left( {p *p_r *p_s } \right) -
H_b \left( {p_s } \right)\right),
\end{align}
where $p_1*p_2$ denotes the discrete convolution operation of two
Bernoulli distributions with parameters $p_1$ and $p_2$, i.e., $p_1
*p_2 = p_1 \left( {1 - p_2 } \right) + p_2 \left( {1 - p_1 }
\right)$, and $H_b\left( p \right) =  - p\log _2 p - \left( {1 - p}
\right)\log _2 \left( {1 - p} \right)$.
\par As a specific numerical example, setting $p = p_r =
0.15$ and $p_s = 0.1$, we have $C_{\rm bin} = 0.4171$. Note that
without state information at the relay, the channel can be
considered as a relay channel with reversely degraded components in
\cite{cover1979capacity}. In this case, the best rate achieved is
given by {\cite[Theorem 2]{cover1979capacity}}:
\begin{align}
 C_{\rm {bin,\;no\;SI}} &= \mathop {\max }\limits_{p\left( x \right)} \mathop {\max
}\limits_{x_R } I\left( {X;Y\left| {X_R  = x_R } \right.} \right) \\
  &= H_b \left( {p*p_s } \right) - H_b \left( {p_s } \right) \\
  &= 0.2912.
\end{align}
Hence $C_{\rm bin} > C_{\rm {bin,\;no\;SI}}$, which assesses the
benefit of state information known at the relay even in a {\it
strictly causal} manner.
\end{example}

\begin{remark}
\par The channel discussed in Example
\ref{example:det:relay}, has a close relationship with the
modulo-additive state-dependent relay model considered by Aleksic,
Razaghi and Yu in \cite{aleksic2009capacity}. Therein, the relay
observes a corrupted version of the noise (state) {\it non-causally}
and has a {\it separate and rate-limited digital link} to
communicate to the destination. For this class of channels, a
compress-and-forward strategy is devised and shown to achieve
capacity. Unlike \cite{aleksic2009capacity}, the relay obtains the
state information {\it noiselessly}, {\it strictly causally} and the
relay-to-destination link is {\it non-orthogonal} to the
source-to-destination link. We have shown in Proposition
\ref{proposition:det:relay} that in this case, the proposed scheme 2
achieves capacity. \remarkend
\end{remark}

\subsection{Message Cooperation Only}
   \par With $C_{RS} = 0$, the model at hand is similar to the one studied in \cite{elgamal2005capacity}, where capacity was obtained for a {\it state-independent} channel in which a general
  noisy channel models the source-to-relay link. For this scenario, the optimal
  coding strategy was found to split the message into
  two parts, one decoded by the relay and sent cooperatively with
  the source to the destination and the other sent directly from the source to the
  destination. By setting $S = V = \emptyset$ and $C_{RS} = 0$ in
  \eqref{equ:achievable:2}, we recover a special case of the capacity obtained in
  \cite{elgamal2005capacity} with noiseless source-to-relay link.
  \par For {\it state-dependent} channels, a general achievable rate can be
  identified through $R_{2}$ in \eqref{equ:achievable:2} by setting $C_{RS} = 0$. Moreover, when the source-to-relay
  capacity is large enough, we are able to characterize the capacity as
  follows. Notice that this capacity result holds for an arbitrary
  $C_{RS}$, not necessary $C_{RS} = 0$.
  \begin{proposition} \label{proposition:messcooponly}
 Let ${\cal P}_{22}^* $ denote the set of distributions defined by:
\begin{align}
 {\cal P}_{22}^*  = \left\{ {p\left({s,x,x_R,y}\right) : p\left(s\right) p\left({x, x_R}\right) p\left({y\left| {s,x,x_R }
\right.}\right) } \right\}. \label{equ:inputd:22}
\end{align}
If $C_{SR}
 \ge \mathop {\max }\limits_{{\cal P}_{22}^* } {I\left(
{X,X_R;Y} \right)} $ and arbitrary $C_{RS}$, the capacity $C_{22}$
is given by:
\begin{align} \label{equ:capacity:22}
C_{22}  = \mathop {\max }\limits_{{\cal P}_{22}^* } {I\left(
{X,X_R;Y} \right)},
\end{align}
and is achieved by message cooperation only.
\end{proposition}
\begin{IEEEproof} When $C_{SR}
 \ge C_{22} $, the source can share a message $w$ of rate $C_{22}$ with the
relay through the conferencing link. By setting $U = X$ and $V =
\emptyset$ in \eqref{equ:achievable:2} and removing redundant
bounds, we establish the achievability part. The converse part
follows directly from \eqref{equ:simple:upperbound}.
\end{IEEEproof}

\begin{remark} \label{remark:ignore:state}
\par The capacity identified above is the same as without any state information at the relay. This result implies that when the relay is
cognizant of the entire message, message transmission always
outperforms sending information about the channel states. In other
words, no benefits can be reaped if the relay allocates part of its
transmission resources to state forwarding. This can be seen as a
consequence of the fact that in a point-to-point channel, no gain is
possible by exploiting availability of strictly causal state
information. \remarkend
\end{remark}

\begin{remark}
\par The capacity result of Proposition \ref{proposition:messcooponly} has been proved by using scheme 2 for achievability. However, it {\it can} also be
obtained with scheme 1 of Proposition \ref{proposition:achievable1}
by setting $U = X$ and $V = \emptyset$. This may not be surprising
since the two schemes differ most notably in the way state
information is processed at encoder and decoder, and the capacity
result of Proposition \ref{proposition:messcooponly} is achieved
with full message cooperation. \remarkend
\end{remark}

\subsection{State Cooperation Only}
   \par If $C_{SR} = 0$, no cooperative message transmission is allowed.
   However, through the conferencing link of capacity $C_{RS}$, cooperative state
   transmission between the relay and source is still feasible. A general achievable rate can be
  identified from $R_{2}$ in \eqref{equ:achievable:2} by setting $C_{SR} = 0$. Specifically, when $C_{RS}$ is large enough,
  we have the following corollary.
 \begin{corollary}\label{corollary:statecoop}
 \par Let ${\cal P}_{23}$ denote the set of
 distributions defined by:
 \begin{align}
 {\cal P}_{23}  = \left\{ {p\left({s,v,x,x_R,y}\right) : p\left(s\right) p\left({v\left|{s,x_R}\right.}\right) p\left({x, x_R}\right) p\left({y\left| {s,x,x_R } \right.}\right) }
\right\}. \label{equ:inputd:23}
\end{align}
If $C_{SR} = 0$ and $C_{RS} \ge \mathop {\max }\limits_{{\cal
P}_{23} } I\left( {X_R ;Y} \right)$, any non-negative rate
  smaller than $R_{23}$ is achievable where
\begin{align} \label{equ:achievable:23}
R_{23}  = \mathop {\max }\limits_{{\cal P}_{23} } \min \left(
{I\left( {X;Y\left| {X_R ,V} \right.} \right),I\left( {X,X_R, V ;Y}
\right) - I\left( {V;S\left| X_R \right.} \right)} \right).
\end{align}
 \end{corollary}

\begin{IEEEproof} By setting $C_{SR} =
 0$ and $U = X_R$ in \eqref{equ:achievable:2}, the set of ${\cal P}_{2}$ is specialized to ${\cal P}_{23}$.
 Fix any input distribution in ${\cal P}_{23}$. The first
 term in the $\min $ function of \eqref{equ:achievable:2} is reduced
 to $I\left( {X;Y\left| {X_R ,V} \right.} \right)$ and the second
 term is reduced to $
I\left( {X,X_R ,V;Y} \right) - I\left( {V;S\left| {X_R } \right.}
\right)$. For the third term, it becomes:
\begin{align}
 &I\left( {X,X_R ,V;Y\left| {X_R } \right.} \right) + C_{RS}  - I\left( {V;S\left| {X_R } \right.} \right) \\
  &= I\left( {X,V;Y\left| {X_R } \right.} \right) + C_{RS}  - I\left( {V;S\left| {X_R } \right.} \right) \\
  &\ge I\left( {X,V;Y\left| {X_R } \right.} \right) + I\left( {X_R ;Y} \right) - I\left( {V;S\left| {X_R } \right.} \right)  \label{equ:achievability:23} \\
  &= I\left( {X,X_R ,V;Y} \right) - I\left( {V;S\left| {X_R }
  \right.}
  \right),
\end{align}
where the inequality \eqref{equ:achievability:23} follows from the
assumption on $C_{RS}$. Notice that the third term cannot be smaller
than the second term, hence it is redundant. Therefore, we establish
the achievable rate given by \eqref{equ:achievable:23}.
\end{IEEEproof}

\par The achievable rate \eqref{equ:achievable:23} coincides with the upper bound \eqref{equ:simple:upperbound} for the special class
of relay channels characterized by
\eqref{det:relay:cond1}$-$\eqref{det:relay:cond2}.
\begin{proposition}\label{proposition:capacity:statecoop}
\par Let ${{\cal P}_{23}^* } = {{\cal P}_{23} }$ as defined by
\eqref{equ:inputd:23}.
 If $C_{SR} = 0$,  $C_{RS} \ge \mathop {\max
}\limits_{{\cal P}_{23}^* } I\left( {X_R ;Y} \right)$, and
\eqref{det:relay:cond1}$-$\eqref{det:relay:cond2} are satisfied for
all distributions in ${{\cal P}_{23}^* }$, then the capacity is
given by:
\begin{align}
C_{23}  = \mathop {\max }\limits_{{\cal P}_{23}^* } \min \left(
{H\left( {Y\left| {X_R ,S} \right.} \right),I\left( {X,X_R ;Y}
\right)} \right). \label{equ:capacity:C23}
\end{align}
\end{proposition}
\begin{proof}
The result follows from Corollary \ref{corollary:statecoop}. For the
achievability, set $V = S$ in \eqref{equ:achievable:23} and apply
assumptions \eqref{det:relay:cond1} and \eqref{det:relay:cond2} to
obtain \eqref{equ:capacity:C23}, by similar steps from
\eqref{equ:capacity:C21:achievable1} to
\eqref{equ:capacity:C21:achievable2}. The upper bounds follow from
\eqref{equ:simple:upperbound} and note that the second bound therein
is reduced to $H\left( {Y\left| {X_R ,S} \right.} \right)$ under
assumption \eqref{det:relay:cond1}.
\end{proof}

\begin{remark}
\par Achievability of the capacity
\eqref{equ:capacity:C23} has been proved above via scheme 2. It {\it
cannot} be attained by scheme 1 because of the additional constraint
required to support the transmission of compressed state information
specified by \eqref{equ:constraint1:th1}.\remarkend
\end{remark}

\begin{remark}
\par Compared to the capacity result provided in Proposition \ref{proposition:det:relay} for the same class of channels
\eqref{det:relay:cond1}$-$\eqref{det:relay:cond2}, $C_{23}$ is
potentially larger because a general input distribution is
admissible instead of the product input distribution due to state
cooperation. The resulting cooperative gain will be further
discussed for the Gaussian model in Section
\ref{sec:relay:Gaussian:numerical}.\remarkend
\end{remark}

\begin{remark}
The capacity result of Proposition
\ref{proposition:capacity:statecoop} is derived for $C_{SR} = 0$ and
is thus achieved by state cooperation only. Optimality of state
cooperation only can also be concluded in some case when $C_{SR} > 0
$ and thus message cooperation is possible. For instance, assume
that $H\left(Y\left|\right.X_R, S\right) \ge I\left(X,X_R;Y\right)$
for the distribution in ${{\cal P}_{23}^* }$ that maximizes
\eqref{equ:capacity:C23}. Then it can be proved, following the same
bounds used in Proposition \ref{proposition:capacity:statecoop},
that the capacity of channels satisfying
\eqref{det:relay:cond1}$-$\eqref{det:relay:cond2} is given by
\eqref{equ:capacity:22} and is achieved by state cooperation only.
An instance of this scenario will be considered in Corollary
\ref{corollary:Gaussian:capacity}.  \remarkend
\end{remark}

\section{Cooperation Strategies With Total Conferencing Capacity Fixed}\label{sec:relay:resource:fixed}
\par In the previous sections, we have studied system performance for
given values of the link capacities $C_{SR}$ and $C_{RS}$. Here, we
briefly investigate the optimal capacity allocation between the
source-to-relay and relay-to-source links where the total
conferencing capacity is instead fixed as $ C_{SR} + C_{RS} =
C_{sum}$. In particular, we compare the rates achievable when the
entire capacity is allocated to message cooperation only
$\left(C_{SR} = C_{sum}\;{\rm and}\;C_{RS} = 0\right)$, to state
cooperation only $\left(C_{SR} = 0 \;{\rm and}\;C_{RS} = C_{sum}
\right)$, or to a combination of message and state cooperation
$\left(C_{SR} = C_{opt}
> 0\;{\rm and}\;C_{RS} = C_{sum} - C_{opt}\right)$. We refer the achievable
rates corresponding in the three cases above by scheme 2 as
$R_{2,M}$, $R_{2,S}$ and $R_{2,MS}$ respectively.

\begin{proposition}
\par For scheme 2, when $ C_{SR} + C_{RS} = C_{sum}$ is fixed, we have
\begin{align}
R_{2,S} \le R_{2,MS} \le R_{2,M}.
\end{align}
\end{proposition}
\begin{IEEEproof}
Fix any input distribution of the form \eqref{equ:inputd:2} in $R_2$
of \eqref{equ:achievable:2}. The second term in the $\min$ function
of \eqref{equ:achievable:2} is independent of both $C_{SR}$ and
$C_{RS}$, hence it is independent of $C_{sum}$. The third term, when
$C_{SR}+ C_{RS} = C_{sum}$ is fixed, is the same no matter how one
allocates $C_{sum}$ between $C_{SR}$ and $C_{RS}$. Finally, the
first term increases with $C_{SR}$. Since $C_{SR}$ cannot be greater
than $C_{sum}$, it is optimal to set $C_{SR} = C_{sum}$. It follows
that $R_{2,S} \le R_{2,MS} \le R_{2,M}$.
\end{IEEEproof}

\begin{remark}
For scheme 2, it is optimal to allocate all conferencing resources
for message forwarding, thereby leading to message cooperation only.
In other words, state cooperation is generally not advantageous when
utilizing this scheme if one can arbitrarily allocate the overall
conferencing capacity. Notice that this may not be always possible,
as for instance, in applications where the two conferencing links
are unidirectional channels with fixed capacity, e.g., cables.
Assessing a similar conclusion holds for scheme 1 seems to be more
difficult and is left as an open problem. \remarkend
\end{remark}

\section{Gaussian Model}\label{sec:relay:Gaussian:numerical}
\par In this section, we study the Gaussian model depicted in
Fig.~\ref{fig:model:relay}, in which the destination output $Y_i$ at
time instant $i$ is related to the channel input $X_i$ from the
source, $X_{R,i}$ from the relay, and the channel state $S_i$ as
\begin{align}
Y_i  = X_i  + X_{R,i}  + S_i  + Z_i, \label{equ:Gaussian:model}
\end{align}
where $S_i \sim {\cal N}\left( {0,P_S } \right)$ and $Z_i \sim {\cal
N}\left( {0,N_0} \right)$, are i.i.d., mutually independent
sequences. The channel inputs from the source and relay satisfy the
following average power constraints
\begin{align}
\frac{1}{n}\sum\limits_{i = 1}^n {{\mathbb E}\left[ {X_i^2 }
\right]}  \le P, \;\;\frac{1}{n}\sum\limits_{i = 1}^n {{\mathbb
E}\left[ {X_{R,i}^2 } \right]}  \le P_R.  \label{equ:Gaussian:power}
\end{align}
The conferencing operations, encoding and decoding functions are
defined as in Definition \ref{def:randomcode} except that the
codewords are required to guarantee the input power constraints
\eqref{equ:Gaussian:power}.

\subsection{Achievable Rate}
\par First, we extend the rate
\eqref{equ:achievable:1} achievable by scheme 1 to the Gaussian
model of \eqref{equ:Gaussian:model}$-$\eqref{equ:Gaussian:power}.

\begin{proposition}\label{proposition:Gaussian:scheme1}
For the Gaussian relay channel considered, scheme 1 achieves any
non-negative rate smaller than $R_{1}^{G}$ where
\begin{align}
R_1^{G}  = \mathop {\max }\limits_{\scriptstyle 0 \le \alpha  \le 1
\hfill \atop
  {\scriptstyle 0 \le \beta  \le 1 \hfill \atop
  \scriptstyle \sigma \le P_{Q}  \hfill}} \min \left(
  A_1, A_2, A_3\right) \label{equ:achievable1:Gaussian}
\end{align}
with
\begin{align}
&A_1  = {\cal C}\left( {\frac{{\left( {1 - \alpha } \right)P}}{{N_0
+ \frac{{P_S P_Q }}{{P_S  + P_Q }}}}} \right) +
C_{SR},\\
&A_2  = {\cal C}\left( {\frac{{P + P_R  + 2\sqrt {\alpha \beta PP_R
} }}{{N_0 + \frac{{P_S P_Q }}{{P_S  + P_Q }} }}} \right) - {\cal
C}\left( {\frac{{\left( {P + P_R  + 2\sqrt {\alpha \beta PP_R }  +
N_0 } \right)P_S }}{{\left( {P + P_R  + 2\sqrt {\alpha \beta PP_R }
+ P_S  + N_0 } \right)P_Q }}} \right),\\
&A_3 = {\cal C}\left( {\frac{{\left( {1 - \alpha } \right)P + \left(
{1 - \beta } \right)P_R }}{{N_0 + \frac{{P_S P_Q }}{{P_S  + P_Q }}
}}} \right)- {\cal C}\left( {\frac{{\left( {P + P_R  + 2\sqrt
{\alpha \beta PP_R }  + N_0 } \right)P_S }}{{\left( {P + P_R  +
2\sqrt {\alpha
\beta PP_R }+ P_S  + N_0 } \right)P_Q }}} \right)\nonumber \\
  &\;\;\;\;+ C_{SR}  + C_{RS} \label{equ:Gaussian:rate:a3}
\end{align}
\end{proposition}
where $\alpha$, $\beta$ are the power allocation coefficients at the
source and relay respectively, $P_{Q}$ is the variance of
compression noise selected at the relay and $\sigma$ is a threshold
defined as:
\begin{align}
\sigma  = \frac{{P_S \left( {P + P_R  + 2\sqrt {\alpha \beta PP_R }
+ N_0 } \right)}}{{\left( {P + P_R  + 2\sqrt {\alpha \beta PP_R }  +
P_S  + N_0 } \right)\min \left( {2^{2C_{RS} } \left( {1 +
\frac{{\left( {1 - \beta } \right)P_R }}{{P_S  + N_0 }}} \right) -
1,\frac{{P + P_R  + 2\sqrt {\alpha \beta PP_R } }}{{P_S  + N_0 }}}
\right)}}. \label{equ:achievable1:Gaussian:constraint}
\end{align}

\begin{IEEEproof} [Sketch of Proof] The result follows from \eqref{equ:achievable:1}$-$\eqref{equ:constraint1:th1} by choosing
Gaussian input signals satisfying the power constraints. Explicitly,
the signals are generated as follows. First, choose $ U \sim {\cal
N}\left( {0,1} \right)$. Then, consider $X = \sqrt {\alpha P} U +
\tilde X $, where $ 0 \le \alpha \le 1$ and $\tilde X \sim {\cal
N}\left( {0,\left( {1 - \alpha } \right)P} \right) $, independent of
$U$. Hence, $X \sim {\cal N}\left( {0,P} \right) $. Similarly, set
$X_R = \sqrt {\beta P_R} U + \tilde X_R $, where $ 0 \le \beta  \le
1$ and $\tilde X_R \sim {\cal N}\left( {0,\left( {1 - \beta }
\right)P_R} \right) $, independent of $U$ and ${\tilde X}$. Hence
$X_R \sim {\cal N}\left( {0,P_R} \right) $ and $ {\mathbb E}\left[
{XX_R } \right] = \sqrt {\alpha \beta PP_R }$. Next, set $V = S + Q$
with compression noise $  Q \sim {\cal N}\left( {0,P_{ Q}} \right)$
for some $P_{Q} \ge \sigma$. By standard techniques as in
{\cite[Ch.8 and 9]{info_cover}}, each mutual information term in
\eqref{equ:achievable:1} and \eqref{equ:constraint1:th1} can be
explicitly evaluated, establishing the achievable rate given from
\eqref{equ:achievable1:Gaussian} to
\eqref{equ:achievable1:Gaussian:constraint}.
\end{IEEEproof}

\par Next, we extend the rate
\eqref{equ:achievable:2} achievable by scheme 2 to the Gaussian
model of \eqref{equ:Gaussian:model}$-$\eqref{equ:Gaussian:power}.

\begin{proposition} \label{proposition:Gaussian:scheme2}
For the Gaussian relay channel considered, scheme 2 achieves any
non-negative rate smaller than $R_{2}^{G}$ where
\begin{align} \label{equ:achievable:Gaussian}
R_2^{G}  = \mathop {\max }\limits_{\scriptstyle 0 \le \alpha  \le 1
\hfill \atop
  {\scriptstyle 0 \le \beta  \le 1 \hfill \atop
  \scriptstyle 0 \le P_{Q}  \hfill}} \min \left(
  B_1, B_2, B_3\right)
\end{align}
with
\begin{align}
&B_1  =  {\cal C}\left( {\frac{{\left( {1 - \alpha } \right)P}}{{N_0
+ \frac{{P_S P_Q }}{{P_S  + P_Q }}}}} \right) +
C_{SR},\\
&B_2  = \frac{1}{2}\log _2 \frac{{P + P_R  + 2\sqrt {\alpha \beta
PP_R }  + P_S  + N_0 }}{{N_0  + \frac{{P_S P_Q }}{{P_S  + P_Q }}}} -
{\cal C}\left( {\frac{{P_S }}{{P_Q }}} \right),\\
&B_3  = \frac{1}{2}\log _2 \frac{{\left( {1 - \alpha } \right)P +
\left( {1 - \beta } \right)P_R  + P_S  + N_0 }}{{N_0  + \frac{{P_S
P_Q }}{{P_S  + P_Q }}}} - {\cal C}\left( {\frac{{P_S }}{{P_Q }}}
\right)+ C_{SR}  + C_{RS} \label{equ:Gaussian:rate:b3}
\end{align}
\end{proposition}
where $\alpha$, $\beta$ are the power allocation coefficients at the
source and relay respectively and $P_{Q}$ is the variance of
compression noise selected at the relay.
\begin{IEEEproof}[Sketch of Proof]
We use the same variable definitions as in the proof of Proposition
\ref{proposition:Gaussian:scheme1} with exception that $P_Q$ only
needs to satisfy $P_Q \ge 0$. Then we can explicitly evaluate each
mutual information term in \eqref{equ:achievable:2} following
standard techniques in {\cite[Ch.8 and 9]{info_cover}}. Details are
omitted here for the sake of conciseness.
\end{IEEEproof}

\begin{remark} \label{remark:ignore:state}
\par If the relay ignores the available state
information, it only cooperates with the source in sending the
message information and does not employ the relay-to-source
conferencing link. An achievable rate corresponding to this
situation can be found from
\eqref{equ:achievable:Gaussian}$-$\eqref{equ:Gaussian:rate:b3} by
setting $P_Q \to \infty$, i.e., an infinite variance for the
compression of the state information, and $\beta = 1$, i.e., the
relay allocates all its power to message transmission. We thus
obtain
\begin{align} \label{equ:ignore:state}
R_{{\rm no\;SI}}^{G}  = \mathop {\max }\limits_{0 \le \alpha  \le 1}
\min \left( \begin{array}{l}
 {\cal C}\left( {\frac{{\left( {1 - \alpha } \right)P}}{{N_0  + P_S }}}
\right) + C_{SR} , \\
{\cal C}\left( {\frac{{P + P_R  + 2\sqrt {\alpha PP_R } }}{{N_0  +
P_S }}}
\right)\\
 \end{array} \right).
\end{align}
Notice that the rate is clearly independent of $C_{RS}$. This rate
will be later used for performance comparison. \remarkend
\end{remark}

\subsection{Special Cases and Capacity Results}
\par Now we focus on the special case where $N_0 = 0$ for the Gaussian model of
\eqref{equ:Gaussian:model}$-$\eqref{equ:Gaussian:power}. We first
consider the case with no both message and state cooperation.

\begin{corollary}
If $N_0 = 0$ and the conferencing links satisfy $C_{SR} = C_{RS} =
0$, the capacity is given by:
\begin{align}
C_{\rm no\;coop}^G  = {\cal C}\left( {\frac{{P + P_R }}{{P_S }}}
\right). \label{equ:Gaussian:nocoop}
\end{align}
\end{corollary}
\begin{IEEEproof}
 Notice that the channel discussed here satisfies assumptions \eqref{det:relay:cond1}$-$\eqref{det:relay:cond2}
 in Proposition \ref{proposition:det:relay}. Hence, by extending the results therein to
 continuous alphabets and evaluating each term by the maximum
 entropy theorem \cite{info_cover}, one can obtain the result claimed in this corollary. Note that when providing both
 $S$ and $X_R$ to the destination, the channel from source to
 destination is noiseless and hence the first bound in the $\min$
 function of \eqref{equ:capacity:C21} goes to infinity, and is thus redundant.
\end{IEEEproof}
\begin{remark} \label{remark:state:feedback:coding}
The capacity result indicates that strictly causal state information
at the relay can provide power gain for the channel considered, even
though the relay knows nothing about the message information
intended for destination from the source. In fact, when $N_0 = 0$,
conveying state information from the relay to destination can be
considered as equivalently sending part of message for the source,
as previously discussed in Remark \ref{remark:det:interpretation}.
\par To
elaborate on this insight further, we sketch an alternative
achievable scheme in which we explicitly split the message $W$ from
the source into two parts, $W = \left(W_{s1}, W_{s2}\right)$, with
$W_{s1} \in \left[1: 2^{nR_{s1}}\right]$ and $W_{s2} \in \left[1:
2^{n{R_{s2}}}\right]$. We divide interval $\left[-1, 1\right]$ into
$2^{nR_{s1}}$ subintervals of equal length and map $W_{s1}$ to the
middle points, denoted by $\theta\left(W_{s1}\right)$, of those
subintervals. In addition, we generate $2^{nR_{2s}}$ i.i.d.
sequences $x^n$ with each component satisfying $x_i \sim {\cal
N}\left( {0,P} \right)$, and map $W_{s2}$ to the sequences generated
as $x^n\left(W_{s2}\right)$. Assume that the source wishes to send
$\left(w_{s1}, w_{s2}\right)$ to the destination. The communication
happens in $\left(n+1\right)$ channel uses as follows. In the first
channel use, the source sends out the middle point $\theta \left(
{w_{s1} } \right)$ corresponding to message $w_{s1}$ while the relay
sends $x_{R,1} = 0$. For the remaining $n$ channel uses, the source
sends out each component of $x^n\left(w_{s2}\right)$ in order.
While, for the relay, in the second channel use, it sends out a
scaled version of the state of the previous channel use such that
the power constraint is satisfied at the relay, i.e., $x_{R,2} = \mu
_2 s_1$, where $ \mu _2$ is chosen such that $ x_{R,2} \sim {\cal
N}\left( {0,P_R } \right)$; For $i \ge 3$ channel uses, the relay
sequentially forms the minimum mean squared error (MMSE) estimate $
{{\mathbb E}\left[ {s_1 \left| {{{\tilde y}}_2^{i - 1} } \right.}
\right]}$ with each $ \tilde y_k  = x_{R,k}  + s_k$, $\forall\:k =
2,...,i-1$, based on the available states $s^{i-1}$ and sends out
$x_{R,i}  = u_i \left( {s_1 - {\mathbb E}\left[ {s_1 \left|
{{{\tilde y}}_2^{i - 1} } \right.} \right]} \right)$, where  $ \mu
_i$ is chosen such that $ x_{R,i} \sim {\cal N}\left( {0,P_R }
\right)$. This way, at the end of transmission, the destination
first decodes message $w_{s2}$ by treating the states and
information sent by the relay as noise. Hence, as long as $R_{s2}
\le {\cal C}\left( {\frac{P}{{P_R + P_S }}} \right)$, $w_{s2}$ can
be successfully recovered as $n \to \infty$. After decoding
$w_{s2}$, subtracting $x^n\left(w_{s2}\right)$ from the received
signal, similar to the analysis of the feedback strategy for
point-to-point additive Gaussian channels in
\cite{SKfeedback,YKGamalLecture}, one can show that $w_{s1}$ can be
successfully decoded at rate $R_{s1} = {\cal C}\left( {\frac{{P_R
}}{{P_S }}} \right)$ by the state refinement transmission from the
relay as $n \to \infty$. Overall, rate $R_{s1} + R_{s2} = {\cal
C}\left( {\frac{{P_R }}{{P_S }}} \right) + {\cal C}\left(
{\frac{P}{{P_R + P_S }}} \right) = {\cal C}\left( {\frac{{P + P_R
}}{{P_S }}} \right) = C_{\rm no\;coop}^G $ is thus achieved for the
source. It is noted that a similar feedback coding scheme can be
found in \cite{LS_IZS2010} to achieve the maximum rate for each user
in a two-user MAC with common state information.
 \remarkend
\end{remark}
\par Next, we consider the optimality of state and message cooperation
only following Proposition \ref{proposition:messcooponly} and
\ref{proposition:capacity:statecoop}.

\begin{corollary}\label{corollary:Gaussian:capacity}
\par If $N_0 = 0$ and the conferencing links
satisfy $C_{RS} \ge {\cal C}\left( {\frac{{P + P_R  + 2\sqrt {PP_R }
}}{{P_S }}} \right)$ with arbitrary $C_{SR}$, the capacity is given
by:
\begin{align} \label{Gaussian:capacity}
 C^G = {\cal C}\left( {\frac{{P + P_R  + 2\sqrt {PP_R }
}}{{P_S }}} \right),
\end{align}
and is achieved by state cooperation only. Moreover, if $N_0 = 0$
and the conferencing links satisfy $C_{SR} \ge C^G$ with arbitrary
$C_{RS}$, the capacity is also given by \eqref{Gaussian:capacity},
and is attained by message cooperation only.
\end{corollary}
\begin{remark}
 \par Example 1 in \cite{LS_IZS2010} implies that, if the source knows the state sequence as well, then
 the maximum rate is given by \eqref{Gaussian:capacity}. Corollary \ref{corollary:Gaussian:capacity} then quantifies the minimum capacity
 $C_{RS}$ necessary for this result to be attained on the relay channel of Fig.~\ref{fig:model:relay} where the source is not given the state sequence. \remarkend
\end{remark}

\begin{IEEEproof}
 To prove achievability for the case when $C_{RS}\ge C^G$, we consider a scheme that uses only the
 relay-to-source conferencing link and perform no message
 cooperation so that we can equivalently set $C_{SR} = 0$. Then, we
 can identify the result from Proposition
 \ref{proposition:capacity:statecoop} by simple extension to
 continuous alphabets and maximizing each term by the maximum
 entropy theorem \cite{info_cover}. Alternatively,
 considering the achievable rate \eqref{equ:achievable:Gaussian}
by scheme 2 and setting $N_0 = 0, C_{SR} = 0$, we rewrite $B_1$ to
$B_3$ in the $\min$ function as follows:
\begin{align}
 &B'_1  = {\cal C}\left( {\frac{{\left( {1 - \alpha } \right)P}}{{\frac{{P_S P_Q
}}{{P_S  + P_Q }}}}} \right), \\
 &B'_2  = {\cal C}\left( {\frac{{P + P_R  + 2\sqrt {\alpha \beta PP_R } }}{{P_S }}}
\right), \\
&B'_3 = {\cal C}\left( {\frac{{\left( {1 - \alpha } \right)P +
\left( {1 - \beta } \right)P_R }}{{P_S }}} \right) + C_{RS}.
\end{align}
Further, setting $\alpha \to 1$, $\beta \to 1$ and $P_Q \to 0$ such
that $B'_1 \to \infty$, and under the assumption that $C_{RS} \ge
{\cal C}\left( {\frac{{P + P_R + 2\sqrt {PP_R } }}{{P_S }}}
\right)$, we thus get $ C^G = B'_2 = {\cal C}\left( {\frac{{P + P_R
+ 2\sqrt {PP_R } }}{{P_S }}} \right)$. For the converse part, the
upper bound \eqref{equ:simple:upperbound} reduces to $C^G$ following
from the maximum entropy theorem \cite{info_cover}.
\par Turning to the case when $C_{SR}\ge C^G$, for the achievable
scheme, the relay simply ignores the state information, so that one
can equivalently set $C_{RS} = 0$, and fully cooperates with the
source to transmit the message, so that one achieves rate
\eqref{equ:ignore:state} with $\alpha = 1$, which reduces to
\eqref{Gaussian:capacity} under the given condition for $C_{SR}$.
\end{IEEEproof}
\par From Corollary \ref{corollary:Gaussian:capacity}, we
immediately have the following.
\begin{corollary}
If $N_0 = 0$, and both $C_{RS}$ and $C_{SR}$ are large enough, both
state and message cooperation only are optimal and achieve the full
cooperation bound \eqref{Gaussian:capacity}. Compared to the case
without any cooperation of \eqref{equ:Gaussian:nocoop}, they both
provide cooperative gain.
\end{corollary}

\begin{remark}\label{remark:scheme1:cannot:beat:scheme2}
If $C_{SR}$ is large enough, e.g., $C_{SR} \ge C^G$, scheme 1 can
also achieve capacity, which is attained by setting $P_Q \to \infty$
and $\beta = 1$ in \eqref{equ:achievable1:Gaussian} similar to
Remark \ref{remark:ignore:state}. However, no matter how large
$C_{RS}$ is, scheme 1 cannot achieve capacity if $C_{SR} = 0$. This
can be argued by considering the extreme case with $C_{RS} \to
\infty$. Examining rate
\eqref{equ:achievable1:Gaussian}$-$\eqref{equ:achievable1:Gaussian:constraint}
of scheme 1, we notice that the third term in the $\min$ function is
redundant due to $C_{RS} \to \infty$. With $N_0 = 0$ and $C_{SR} =
0$, the first two terms can be instead rewritten as
\begin{align}
 &A'_1  = {\cal C}\left( {\frac{{\left( {1 - \alpha } \right)P}}{{\frac{{P_S P_Q
}}{{P_S  + P_Q }}}}} \right), \\
 &A'_2  = {\cal C}\left( {\frac{{P + P_R  + 2\sqrt {\alpha \beta PP_R } }}{{P_S }}}
\right),
\end{align}
along with an additional constraint
\begin{align}
P_Q \ge \sigma  = \frac{{P_S^2 }}{{P + P_R  + 2\sqrt {\alpha \beta
PP_R }  + P_S }}.
\end{align}
To achieve capacity $C^G$, we need to set $\alpha \to 1$, $\beta \to
1$ and $P_Q \to 0$ as discussed in Corollary
\ref{corollary:Gaussian:capacity}. But, notice that $P_Q$ is always
bounded below by a nonzero threshold, which implies that $P_Q \to 0$
cannot be satisfied. Therefore, it can be concluded that scheme 1
cannot achieve capacity by state cooperation only no matter how
large $C_{RS}$ is. Recall that in scheme 1, the additional
constraint comes from the fact that the destination needs to decode
the compressed state explicitly, as discussed in Remark
\ref{remark:achievable1}. Compared to this scheme, the advantages of
scheme 2 come from joint decoding of message and compression
indices. \remarkend
\end{remark}

\subsection{Numerical Results and Discussions}
\par
\begin{figure}[t]
\centering
\includegraphics[width=0.6\textwidth]{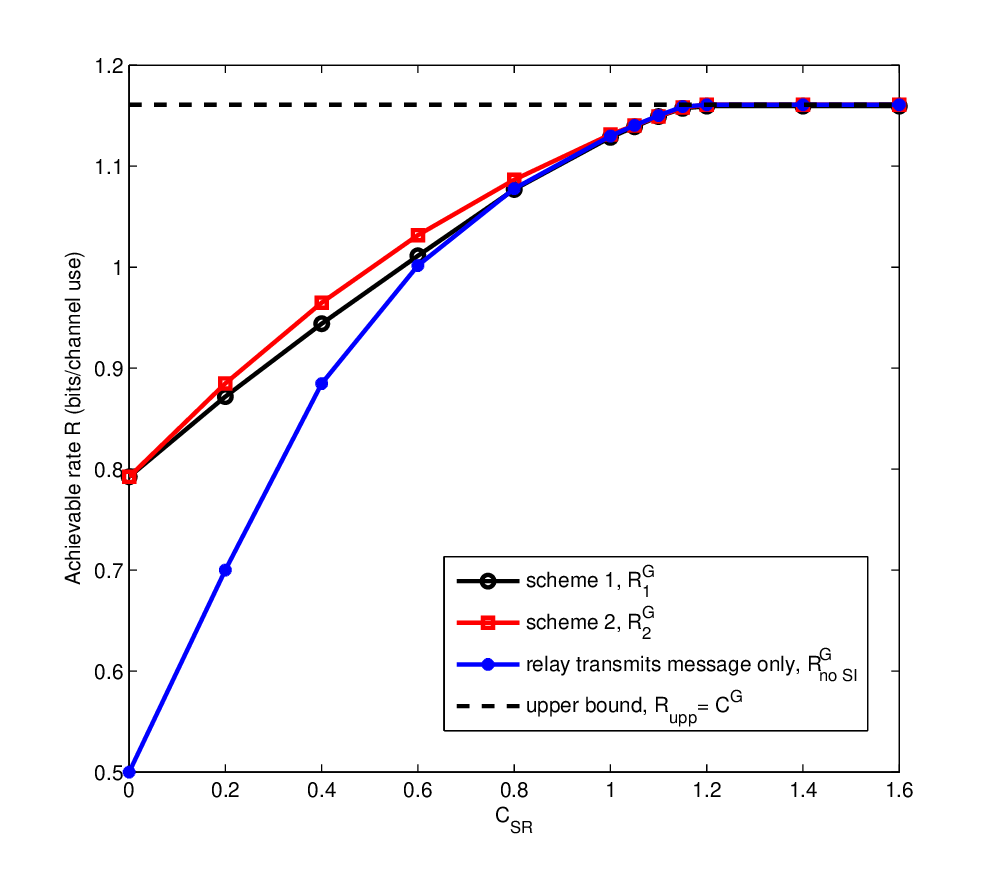} \caption{Comparison of achievable rates between scheme 1 and scheme 2 for message cooperation only $\big(C_{RS} = 0, P = P_R = P_S =
1, N_0 = 0\big)$.}\label{fig:Noiseless:MOnly}
\end{figure}

 We now present some numerical results. We start from the special case with $N_0 = 0$ studied in Corollary \ref{corollary:Gaussian:capacity}.
We first compare the performance of scheme 1 and scheme 2 for
message cooperation only, i.e., $C_{RS} = 0$. In
Fig.~\ref{fig:Noiseless:MOnly}, we plot the achievable rates versus
conferencing capacity $C_{SR}$. We also plot the rate $R_{\rm no
\;SI}^{G}$ in  \eqref{equ:ignore:state} that is achieved when the
relay does not use the available side information. It can be seen
that scheme 2 outperforms scheme 1 in general, consistently with
Proposition \ref{proposition:scheme2:beat:scheme1}. Moreover, if
$C_{SR}$ is large enough, both schemes achieve the upper bound
\eqref{Gaussian:capacity} and the optimal strategy is to let the
relay ignore the state information as provided in Corollary
\ref{corollary:Gaussian:capacity}. But this strategy is suboptimal
for smaller $C_{SR}$. The benefits of state transmission from the
relay to the destination are thus clear from this example.

\par
\begin{figure}[t]
\centering
\includegraphics[width=0.6\textwidth]{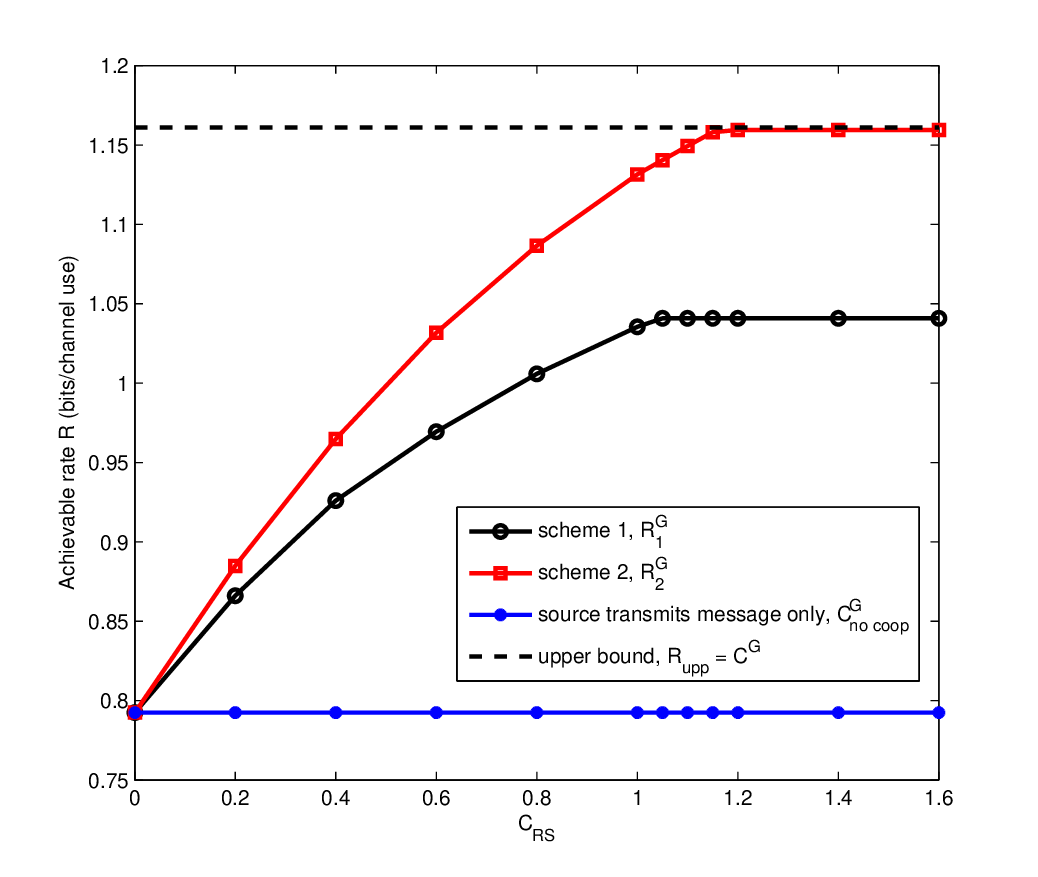} \caption{Comparison of achievable rates between scheme 1 and scheme 2 for state cooperation only $\big(C_{SR} = 0, P = P_R = P_S = 1
, N_0 = 0\big)$.}\label{fig:Noiseless:SOnly}
\end{figure}

 Next, we
consider state cooperation only, that is, $C_{SR} = 0$, and compare
the achievable rates for two schemes in
Fig.~\ref{fig:Noiseless:SOnly} with the upper bound
\eqref{Gaussian:capacity}. We also plot the achievable rate $C_{\rm
no\;coop}^G$ in \eqref{equ:Gaussian:nocoop} that is attained when
the source transmits message only. The benefits of cooperative state
transmission by the source are clear from the figure. Moreover, if
$C_{RS}$ is large enough, scheme 2 is seen to achieve the upper
bound, as proved in Corollary \ref{corollary:Gaussian:capacity}.
Instead, scheme 1 cannot, as discussed in Remark
\ref{remark:scheme1:cannot:beat:scheme2}.

\par We now get further insights into system performance by letting
$N_0 \ne 0$. We set $P = P_R = P_S = 1$ and vary $N_0$ such that the
resulting signal-to-noise ratio, or interfered state-to-noise ratio,
$\gamma  = 10\log _{10} \left( {1/N_0 } \right)$ lies between $
\left[ { - 5:30} \right]$ dB.

\par
\begin{figure}[t]
\centering
\includegraphics[width=0.52\textwidth]{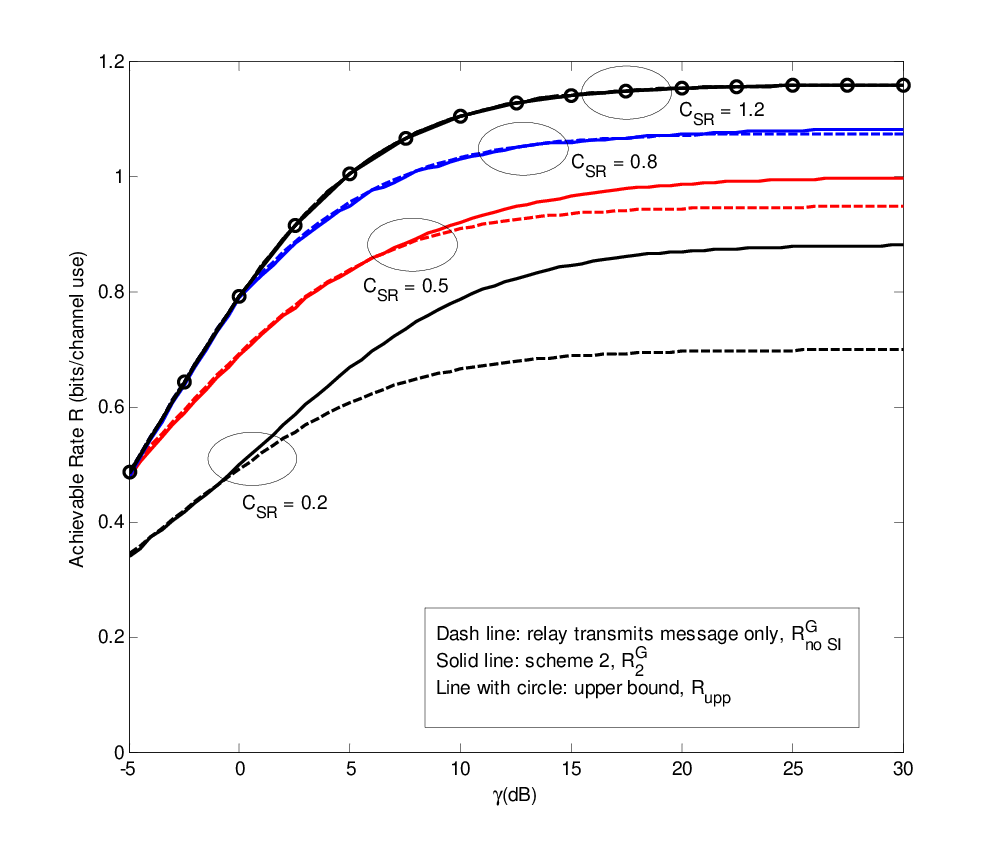}
\caption{Comparison of achievable rates for message cooperation only
by scheme 2 $ \big(C_{SR} = \left\{ {0.2,0.5,0.8,1.2} \right\},
C_{RS} = 0, P = P_R = P_S = 1, \gamma =
10\log_{10}\left(1/N_0\right)\left(dB\right) \big)$.
}\label{fig:messcooponly}
\end{figure}

 We focus on scheme 2 and consider message cooperation only, i.e., $C_{RS} = 0$.
Fig.~\ref{fig:messcooponly} shows the rates achievable by scheme 2
and by the same scheme when the relay ignores the state information
\eqref{equ:ignore:state} versus $\gamma$. It can be seen that in
general state transmission from the relay can provide rate
improvement, as also shown in Fig.~\ref{fig:Noiseless:MOnly}. With
$C_{SR}$ increasing, the achievable rate increases until it
saturates at the upper bound \eqref{equ:simple:upperbound} when
$C_{SR}$ is large enough. For example, as shown in
Fig.~\ref{fig:messcooponly}, when $C_{SR} = 1.2$, the achievable
rate overlaps with the upper bound.

\par
\begin{figure}[t]
\centering
\includegraphics[width=0.51\textwidth]{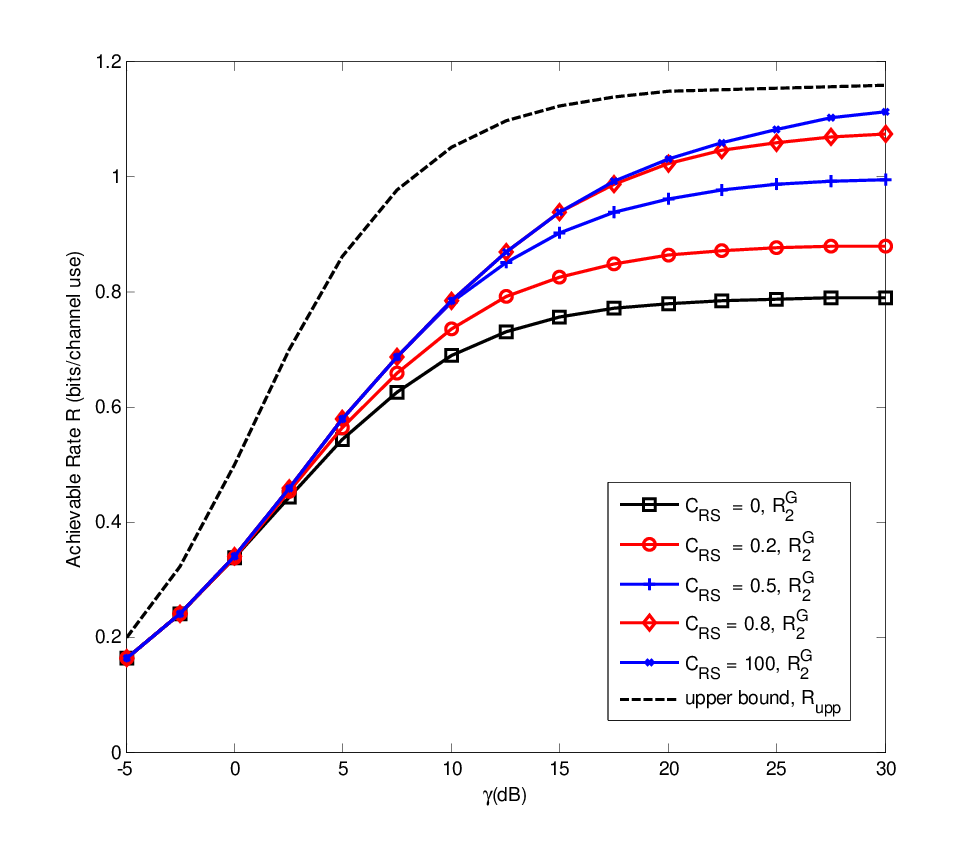} \caption{Comparison of achievable
rates for state cooperation only by scheme 2 $\big(C_{SR} = 0,
C_{RS} = \left\{ {0,0.2,0.5,0.8,100} \right\}, P = P_R = P_S = 1,
\gamma = 10\log_{10}\left(1/N_0\right) \left(dB\right) \big)$.
}\label{fig:statecooponly}
\end{figure}

 We now consider state cooperation only, that is, $C_{SR} = 0$. Fig.~\ref{fig:statecooponly}
shows the rate achievable by scheme 2. The upper bound therein also
refers to \eqref{equ:simple:upperbound}. It can be seen that
cooperative state transmission by the source is general
advantageous, as compared to the performance without cooperation,
i.e., $C_{RS} = 0$. However, unlike the case of message cooperation
only, even if $C_{RS}$ is large enough, e.g., $C_{RS} = 100$ in
Fig.~\ref{fig:statecooponly}, the upper bound is not achievable in
general. This is unlike the noiseless case studied in
Fig.~\ref{fig:Noiseless:SOnly}, due to the fact that noise makes the
state information at the destination less valuable (see Remark
\ref{remark:det:interpretation} and
\ref{remark:state:feedback:coding}).

\par
\begin{figure}[t]
\centering
\includegraphics[width=0.55\textwidth]{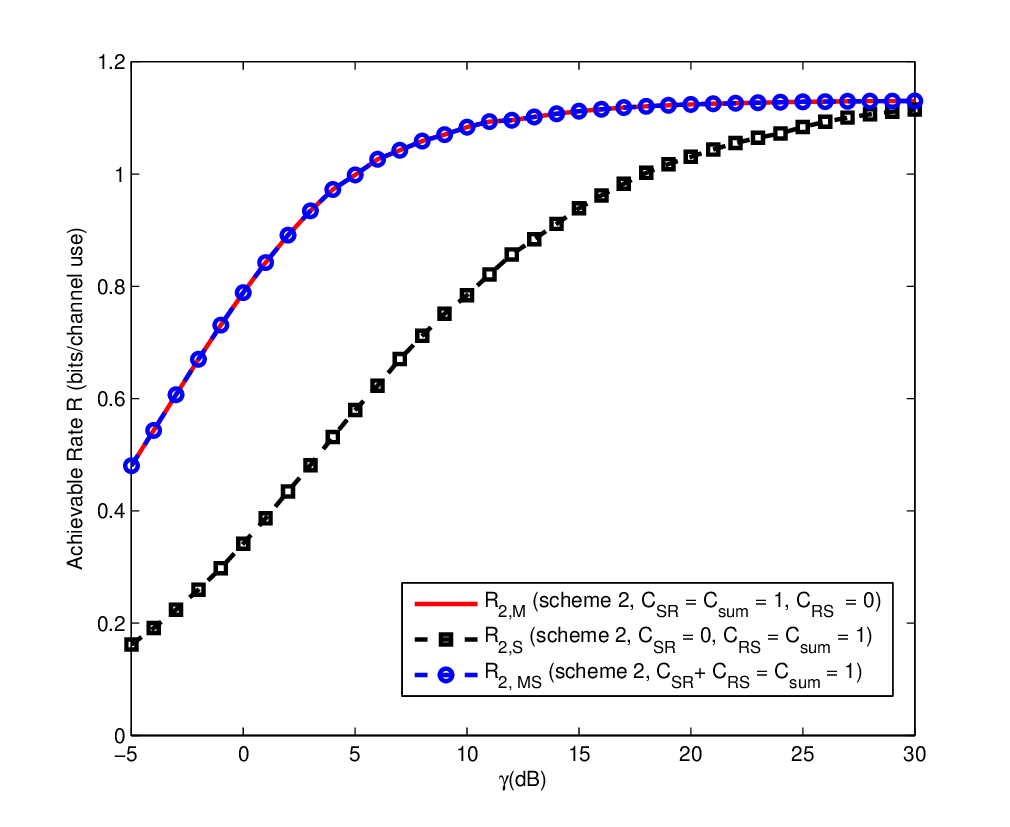} \caption{Comparison of achievable rates for
different cooperation strategies when the total conferencing
capacity is fixed $\big(C_{sum} = 1, P = P_R = P_S = 1, \gamma =
10\log_{10}\left(1/N_0\right) \left(dB\right)
\big)$.}\label{fig:MessageStateCoopScheme12}
\end{figure}

 Finally, we consider the case when the total conferencing capacity $C_{SR} + C_{RS} = C_{sum}$ is
fixed as discussed in Section \ref{sec:relay:resource:fixed}. Under
this assumption, we have shown in Section
\ref{sec:relay:resource:fixed} that for scheme 2, it is enough to
devote all the capacity for message conferencing, thereby leading to
message cooperation only. We corroborate this analytical result via
a specific example in Fig.~\ref{fig:MessageStateCoopScheme12}. It
can be seen that a combination of both message and state cooperation
is able to provide rate improvements as compared to cooperation on
state only, while message cooperation only is always optimal.

\section{Conclusion}\label{sec:relay:conclusion}
\par In this work, we have focused a state-dependent relay channel where state information is available at the relay
in a strictly causal fashion. Assuming that source and relay can
communicate via conferencing links, cooperation is enabled for both
transmission of message and state information to the destination.
First, we have proposed two coding schemes that exploit both message
and state cooperation. The coding scheme inspired by noisy network
coding outperforms the more conventional strategy based on block
Markov coding and backward decoding. Next, capacity results have
been established for some special cases, including no cooperation,
message cooperation only and state cooperation only for a class of
channels. We have also elaborated on the issue of optimal capacity
allocation between the source-to-relay and relay-to-source
conferencing links. Finally, we have characterized achievable rates
for the Gaussian model and obtained some capacity results. In
general, our results point to the advantage of state information at
the relay, despite it being known only strictly causally. This is
unlike point-to-point channels. Moreover, for given conferencing
capacities, both state and message cooperation can improve the
achievable rate.

 \appendices
\section{Proof of Proposition \ref{proposition:achievable1}}
\label{appendix:proposition:achievable1}
\par Throughout the proof, for a joint probability
distribution $p\left(xy\right)$, the set of $\epsilon$-typical
$n$-sequences according to $p\left(xy\right)$ is denoted by
$A_{\epsilon}^n \left( {XY} \right)$. When the distribution, with
respect to which typical sequences are defined, is clear from the
context, we will use $A_{\epsilon}^n$ for short.
\par Now we present the
achievable scheme. Consider $b$ blocks of transmission. We randomly
and independently generate codebooks for each block.
\begin{itemize}
  \item {\bf Codebook Generation}: \\Fix a joint distribution \\$p\left( {s,v,u,x,x_R ,y} \right)
= p\left(s\right) p\left({v\left| s \right.}\right) p\left(u\right)
p\left({x\left| {u } \right.}\right) p\left({x_R\left|
{u } \right.}\right) p\left({y\left| {s,x,x_R } \right.}\right)$.\\
Define rates $R = R_c  + R_p $ with $ 0 \le R_c \le \min \left(R,
C_{SR}\right)$, and $\tilde R = {\tilde R}_{c} + {\tilde R}_{p} $
with $0 \le {\tilde R}_{c} \le \min \left( {\tilde R,C_{RS} }
\right)$.
\begin{enumerate}
  \item For each block $j$, $j \in \left[ {1:b} \right]$, generate $2^{nR_v }$ i.i.d. sequences $
v_j^n $ according to the marginal probability mass function (PMF) $
p \left( {v_j^n } \right) = \prod\limits_{i = 1}^n {p \left(
{v_{j,i} } \right)}$ for the given $ p \left( {v} \right)$. Index
them as $ v_j^n \left( {l_j } \right)$ with $l_j  \in \left[
{1:2^{nR_v } } \right]$. First partition the set $ \left[ {1:2^{nR_v
} } \right]$ into $2^{n {{\tilde R}_c}}$ superbins of equal size
with each containing $2^{n\left( {R_v  - {\tilde R}_c} \right)} $
$v_j^n\left( {l_j } \right)$ codewords. Then further partition the
codewords in each superbin into $2^{n {{\tilde R}_{p}}}$ bins of
equal size. Then each bin contains $2^{n\left( {R_v - {\tilde R}}
\right)} $ codewords. Index each superbin as
$B_{s,j}\left(t_{c,j}\right)$ while index each bin as
$B_j\left(t_{c, j}, t_{p, j}\right)$ with $t_{c, j} \in \left[
{1:2^{n {\tilde R}_{c} } } \right]$, $t_{p, j} \in \left[ {1:2^{n
{{\tilde R}_{p}}} } \right]$.
  \item For each block $j$, generate $2^{n\left( {R_{c}  + {{\tilde R}_{c}}}
  \right)}$ i.i.d. sequences $u_j^n$ according to $
p \left( {u_j^n } \right) = \prod\limits_{i = 1}^n {p \left(
{u_{j,i} } \right)}$ for the given $p\left(u\right)$. Index them as
$u_j^n \left( {w_{c,j} ,t_{c, j - 1} } \right)$ with $w_{c,j}  \in
\left[ {1:2^{nR_{c} } } \right]$ and $t_{c, j - 1}  \in \left[
{1:2^{n{\tilde R}_{c}} } \right]$.
  \item For each block $j$, for each $u_j^n \left( {w_{c,j} ,t_{c, j
- 1} } \right)$, generate $2^{n{R_{p}}}$
  i.i.d. sequences $x_j^n$ according to the conditional PMF $
p \left( {x_j^n \left| {u_j^n \left( {w_{c,j} ,t_{c,j - 1} }
\right)} \right.} \right) = \prod\limits_{i = 1}^n {p \left(
{x_{j,i} \left| {u_{j,i} \left( {w_{c,j} ,t_{c,j - 1} } \right)}
\right.} \right)}$ for the given $p\left( {x\left| u \right.}
\right)$. Index them as $ x_j^n \left( {w_{p,j} \left| {w_{c,j} ,}
\right.t_{c,j - 1} } \right)$ with $w_{p,j} \in \left[ {1:2^{nR_{p}}
} \right]$.
  \item For each block $j$, for each $u_j^n \left( {w_{c,j} ,t_{c, j
- 1} } \right)$, generate $2^{n{{\tilde R}_{p}}}$
  i.i.d. sequences $x_{R,j}^n$ according to the conditional PMF $
p \left( {x_{R,j}^n \left| {u_j^n \left( {w_{c,j} ,t_{c,j - 1} }
\right)} \right.} \right) = \prod\limits_{i = 1}^n {p \left(
{x_{R,j,i} \left| {u_{j,i} \left( {w_{c,j} ,t_{c,j - 1} } \right)}
\right.} \right)}$ for the given $p\left( {x_R \left| u \right.}
\right)$. Index them as $ x_{R,j}^n \left( {t_{p,j-1} \left|
{w_{c,j} ,} \right.t_{c,j - 1} } \right)$ with $t_{p,j-1} \in \left[
{1:2^{n {\tilde R}_{p}} } \right]$.

\end{enumerate}
  \item {\bf Encoding}:
  \par At the beginning of each block, through conferencing link
  $C_{SR}$, the common message $w_{c,j}$ can be perfectly conveyed to
  the relay as long as $bn{R_c} \le bnC_{SR}$, which implies that
  \begin{align}
   R_c  \le C_{SR}. \label{equ:th1:CSR}
  \end{align}
  Similarly, the superbin index can be delivered to the source as
  long as
  \begin{align}
  {\tilde R}_{c}  \le C_{RS}.\label{equ:th1:CRS}
 \end{align}

  \par Then we have the following encoding operations:
  \begin{enumerate}
    \item $j=1$: To send $
w_1  = \left( {w_{c,1} ,w_{p,1} } \right)$ to the destination, the
source sends out codeword $ x_1^n \left( {w_{p,1} \left| {w_{c,1}
,1} \right.} \right)$ while the relay sends out codeword $x_{R,1}^n
\left( {1 \left| {w_{c,1} ,1} \right.} \right)$.
    \item $j=\left[ {2:b-1} \right]$: Assume $w_j  = \left( {w_{c,j} ,w_{p,j} }
    \right)$ to be sent in the $j$th block. At the end of the $\left(j-1\right)$th block, the relay learns the entire state sequence, i.e.,
    $s_{j - 1}^n$, and looks for an index (compression index) $l_{j-1} $ such
    that $ \left( {s_{j-1}^n ,v_{j-1}^n \left(l_{j-1}\right)}
\right) \in A_\epsilon^n$. If more than one such indices are found,
choose the smallest one. If there is no such an index, choose an
arbitrary index at random from $\left[ {1:2^{nR_{v} } } \right]$.
Let $\left(t_{c,j-1}, t_{p,j-1}\right)$ be the bin index pair
associated with $v_{j-1}^n(l_{j-1})$. Then codeword $ x_j^n \left(
{w_{p,j} \left| {w_{c,j} ,t_{c, j-1}} \right.} \right)$ is sent out
by the source and codeword $x_{R,j}^n \left( {t_{p,j} \left|
{w_{c,j} ,t_{c, j-1}} \right.} \right)$ is sent out by the relay.
    \item $j= b$: No new message is sent at the source. Hence, the
source sends out codeword $ x_b^n \left( {1 \left| {1 ,t_{c, b-1}}
\right.} \right)$ while the relay sends out codeword $ x_{R, b}^n
\left( {t_{p,b-1} \left| {1 ,t_{c, b-1}} \right.} \right)$.
  \end{enumerate}

  \item {\bf Decoding}:
  \par At the end of $b$ blocks of transmission, the destination
  performs backward decoding. It first retrieves the bin index pair $\left(t_{c, b-1}, t_{p, b-1}\right)$
   through reception of $b$th block, then it decodes the
  compression index $l_{b-1}$ by using the received signal $y_{b-1}^n$ and finally it decodes the message $\left(w_{c,b-1},w_{p,b-1}\right)$ for block
  $\left(b-1\right)$ using the compressed state information $v_{b-1}^n\left(l_{b-1}\right)$. This decoding operation is repeated
  for all blocks back to the first.
  \par Specifically, the decoding procedure for message $\left(w_{c,j}, w_{p,j}\right)$ of block $j$ is as follows. Assume that $
\left( {w_{c,j+1} ,w_{p,j+1} } \right)$ are perfectly decoded from
the previous estimate. Now the destination looks for an unique bin
index pair $\left({\hat t}_{c, j}, {\hat t}_{p, j}\right)$ such that
\begin{align}
\left( {x_{j+1}^n \left( {w_{p,j+1} \left| {w_{c,j+1} ,\hat t_{c, j}
} \right.} \right),{x_{R,j+1}^n \left( {\hat t_{p,j} \left|
{w_{c,j+1},\hat t_{c, j}} \right.}
\right)},u_{j+1}^n\left(w_{c,j+1}, {\hat t}_{c, j}\right), y_{j+1}^n
} \right) \in A^n_\epsilon \label{equ:decoding:rule:T}
\end{align}
If there is none or more than one such bin index pairs found, the
destination reports an error. Once it finds such a $\left(t_{c,j},
t_{p,j}\right)$, it looks for an unique compression index $\hat
l_{j}$ such that
\begin{align}
\left( {v_j^n \left( {\hat l_{j} } \right),y_{j}^n } \right) \in
A_\epsilon^n, \label{equ:decoding:rule:V1}
\end{align}
and
\begin{align}
\hat l_{j}  \in B_j\left( {t_{c,j}, t_{p,j} } \right).
\label{equ:decoding:rule:V2}
\end{align}
If there is none or more than one such compression indices found,
the destination reports an error. Once it finds such a $l_{j}$, the
destination looks for an unique message $\hat w_{j} = \left( {\hat
w_{c,j} ,\hat w_{p,j} } \right)$ such that
\begin{align}
\left( {x_j^n \left( {{\hat w}_{p,j} \left| {{\hat w}_{c,j} ,t_{c,j
- 1} } \right.} \right),x_{R,j}^n \left( {{t}_{p,j - 1} \left|
{{\hat w}_{c,j} ,t_{c,j - 1} } \right.} \right), u_j^n\left({\hat
w}_{c,j}, { t}_{c, j-1}\right), v_j^n \left( {l_{j} }
\right),y_{j}^n } \right) \in A_\epsilon^n
\label{equ:decoding:rule:M}
\end{align}
for some $t_{c,j - 1}  \in \left[ {1:2^{n {\tilde R}_{c} } }
\right]$, $t_{p,j - 1}  \in \left[ {1:2^{n {\tilde R}_{p} } }
\right]$.

  \item {\bf Analysis of Probability of Error}:
  \par Let $\Pr\left(E_j\right)$ denote the average probability of error for each block $j$ as defined in \eqref{equ:prob:error}.
  To bound the overall probability of error, say $P_e$, without loss of generality (WLOG),
  assume $\left( {w_{c,j} ,w_{p,j} } \right) = \left(1,1\right) $ are sent for each block $j$.
  Also denote the compression index selected
 by the relay for each block by $L_{j-1}$
 and the corresponding bin index pair for each block by $\left(T_{c,j-1},
 T_{p,j-1}\right)$. Note
 that, following the chain rule,
 \begin{align}
  P_e &= \Pr \left( {\bigcup\limits_{j = 1}^b {E_j } } \right) \\
  &\le \Pr \left( {E_b } \right) + \sum\limits_{j = 1}^{b - 1} {\Pr \left( {E_j \left| {\bigcap\limits_{i = j + 1}^b {E_i^c } } \right.}
  \right)},
 \end{align}
 where since there is no new message sent in the last block, we have $\Pr \left( {E_b } \right)=0$. In the following, we focus on $\Pr \left( {E_j \left|
{\bigcap\limits_{i = j + 1}^b {E_i^c } } \right.} \right)$, i.e.,
the probability of error conditioned on not having errors in block
$j+1,...,b$ for each
 block $j$, $j = 1,...,b - 1$, and we show that $\Pr \left( {E_j \left|
{\bigcap\limits_{i = j + 1}^b {E_i^c } } \right.} \right) \to 0$ as
$n \to \infty$ if conditions \eqref{equ:th1:bound1},
\eqref{equ:th1:bound2}$-$\eqref{equ:th1:bound4}, and
\eqref{equ:th1:bound5}$-$\eqref{equ:th1:bound7} are satisfied.

  \par Define the encoding error event for each block as follows:
  \begin{align}
{E}_{j,0}  = \left\{ {\left( {v_j^n \left( {l_j } \right),s_j^n }
\right) \notin A_{\epsilon}^n ,\forall \:l_j  \in \left[ {1:2^{nR_v
} } \right]} \right\}.
  \end{align}
  \par The error events correspond to decoding $T_{c,j}$ and
  $T_{p,j}$ based on rule \eqref{equ:decoding:rule:T} are given by:
  \begin{align}
E_{j,1}  = E_{j,11}^c  \bigcup E_{j,12} \bigcup E_{j,13}  \bigcup
E_{j,14}
  \end{align}
  with
  \begin{align}
&E_{j,11}  = \left\{ {\left( {x_{j + 1}^n \left( {1\left| {1,T_{c,j}
} \right.} \right),x_{R,j + 1}^n \left( {T_{p,j} \left| {1,T_{c,j} }
\right.} \right),u_{j + 1}^n \left( {1,T_{c,j} }
\right),y_{j + 1}^n } \right) \in A_\epsilon^n } \right\},\\
&E_{j,12}  = \left\{ \begin{array}{l}
 \left( {x_{j + 1}^n \left( {1\left| {1,t_{c,j} } \right.} \right),x_{R,j + 1}^n \left( {T_{p,j} \left| {1,t_{c,j} } \right.} \right),u_{j + 1}^n \left( {1,t_{c,j} } \right),y_{j + 1}^n } \right) \in A_\epsilon^n , \\
 {\;\;\;\;\;\;\;\;\;\;\;\;\;\;\;\;\rm{for}}\:{\rm{some}}\:\:t_{c,j}  \ne T_{c,j}  \\
 \end{array} \right\}, \\
&E_{j,13}  = \left\{ \begin{array}{l}
 \left( {x_{j + 1}^n \left( {1\left| {1,T_{c,j} } \right.} \right),x_{R,j + 1}^n \left( {t_{p,j} \left| {1,T_{c,j} } \right.} \right),u_{j + 1}^n \left( {1,T_{c,j} } \right),y_{j + 1}^n } \right) \in A_{\epsilon}^n , \\
 {\;\;\;\;\;\;\;\;\;\;\;\;\;\;\;\;\rm{for}}\:{\rm{some}}\:\:t_{p,j}  \ne T_{p,j}  \\
 \end{array} \right\}, \\
&E_{j,14}  = \left\{ \begin{array}{c}
 \left( {x_{j+1}^n \left( {1\left| {1,t_{c,j} } \right.} \right),x_{R,j+1}^n \left( {t_{p,j} \left| {1,t_{c,j} } \right.} \right),u_{j+1}^n \left( {1,t_{c,j} } \right),y_{j+1}^n } \right) \in A_{\epsilon}^n , \\
 {\mathrm {for\;some}} \;\;t_{c,j}  \ne T_{c,j}, t_{p,j}  \ne T_{p,j}  \\
 \end{array} \right\}.
  \end{align}
  \par The error events correspond to decoding $L_{j}$ according to
  rule \eqref{equ:decoding:rule:V1}$-$\eqref{equ:decoding:rule:V2}
  are given by:
  \begin{align}
E_{j,2}  = E_{j,21}^c  \bigcup E_{j,22}
  \end{align}
  with
  \begin{align}
&E_{j,21}  = \left\{ {\left( {v_j^n \left( {L_{j} } \right),y_{j}^n } \right) \in A_\epsilon^n } \right\},\\
&E_{j,22}  = \left\{ {\left( {v_j^n \left( {l_{j} } \right),y_{j}^n
} \right) \in A_{\epsilon}^n ,{\mathrm {for\;some}}\;\;l_{j}  \ne
L_{j}, l_{j}\in B_j(T_{c,j}, T_{p,j}) } \right\}.
  \end{align}
  \par The error events correspond to decoding message $\left(w_{c,j},
  w_{p,j}\right)$ according to rule \eqref{equ:decoding:rule:M} are
  given by:
  \begin{align}
E_{j,3}  = E_{j,31}  \bigcup E_{j,32} \bigcup E_{j,33}
  \end{align}
  with
  \begin{align}
 &E_{j,31}  = \left\{ {\begin{array}{*{20}c}
   {\left( \begin{array}{l}
 x_j^n \left( {1\left| {\hat w_{c,j} ,t_{c,j - 1} } \right.} \right),x_{R,j}^n \left( {t_{p,j - 1} \left| {\hat w_{c,j} ,t_{c,j - 1} } \right.} \right), \\
 u_j^n \left( {\hat w_{c,j} ,t_{c,j - 1} } \right),v_j^n \left( {L_j } \right),y_j^n  \\
 \end{array} \right) \in A_\epsilon^n ,}  \\
   {\:\:\:\:\:\:\:\:{\rm{for}}\:{\rm{some}}\:\:\hat w_{c,j}  \ne 1,t_{c,j - 1}  \in \left[ {1:2^{n\tilde R_c } } \right] , t_{p,j - 1}  \in \left[ {1:2^{n\tilde R_p } } \right]}  \\
\end{array}} \right\}, \label{equ:error:event31}\\
 &E_{j,32}  = \left\{ {\begin{array}{*{20}c}
   {\left( \begin{array}{l}
 x_j^n \left( {\hat w_{p,j} \left| {1,t_{c,j - 1} } \right.} \right),x_{R,j}^n \left( {t_{p,j - 1} \left| {1,t_{c,j - 1} } \right.} \right), \\
 u_j^n \left( {1,t_{c,j - 1} } \right),v_j^n \left( {L_j } \right),y_j^n  \\
 \end{array} \right) \in A_\epsilon^n ,}  \\
   {\:\:\:\:\:\:\:\:{\rm{for}}\:{\rm{some}}\:\:\hat w_{p,j}  \ne 1,t_{c,j - 1}  \in \left[ {1:2^{n\tilde R_c } } \right],t_{p,j - 1}  \in \left[ {1:2^{n\tilde R_p } } \right]}  \\
\end{array}} \right\}, \label{equ:error:event32}\\
 &E_{j,33}  = \left\{ \begin{array}{l}
 \left( \begin{array}{l}
 x_j^n \left( {\hat w_{p,j} \left| {\hat w_{c,j} ,t_{c,j - 1} } \right.} \right),x_{R,j}^n \left( {t_{p,j - 1} \left| {\hat w_{c,j} ,t_{c,j - 1} } \right.} \right), \\
 u_j^n \left( {\hat w_{c,j} ,t_{c,j - 1} } \right),v_j^n \left( {L_j } \right),y_j^n  \\
 \end{array} \right) \in A_\epsilon^n ,\: \\
 \:{\rm{for}}\:{\rm{some}}\:\:\hat w_{p,j}  \ne 1,\hat w_{c,j}  \ne 1,t_{c,j - 1}  \in \left[ {1:2^{n\tilde R_c } } \right],t_{p,j - 1}  \in \left[ {1:2^{n\tilde R_p } } \right] \\
 \end{array} \right\}. \label{equ:error:event33}
  \end{align}
  Hence by the union bound,
  \begin{align}
\Pr \left( {E_j \left| {\bigcap\limits_{i = j + 1}^b {E_i^c } }
\right.} \right) \le \Pr \left( {E_{j,0} } \right) + \Pr \left(
{E_{j,1}} \right) + \Pr \left( {E_{j,2} } \right) + \Pr \left(
{E_{j,3} } \right).
  \end{align}
  \begin{enumerate}
    \item By the covering lemma in
    \cite{YKGamalLecture}, $\Pr \left( {E_{j,0} } \right) \to 0$ as long as
    \begin{align}
R_v  > I\left( {V;S} \right) \label{equ:th1:bound1}
    \end{align}
    for sufficiently large $n$.
    \item By the packing lemma in \cite{YKGamalLecture},
    $\Pr \left( {E_{j,1}} \right) \to 0$ as long as
    \begin{align}
 &{\tilde R}_{c}  < I\left( {X,X_R ,U;Y} \right),
 \label{equ:th1:Rc:redudant}\\
 &{\tilde R}_{p}  < I\left( {X_R ;Y\left| {X ,U} \right.} \right), \label{equ:th1:bound2} \\
 &{\tilde R}_{c}  + {\tilde R}_{p}  < I\left( {X,X_R ,U;Y} \right) \label{equ:th1:bound3}
    \end{align}
    for sufficiently large $n$. Note that bound \eqref{equ:th1:bound3} implies
    \eqref{equ:th1:Rc:redudant}, hence \eqref{equ:th1:Rc:redudant}
    is redundant.
    \item $\Pr \left( {E_{j,2} } \right) \to 0$ as long as
    \begin{align}
{R_v - {\tilde R}} < I\left( {V;Y} \right) \label{equ:th1:bound4}
    \end{align}
    for sufficiently large $n$.
    \item For $\Pr
\left( {E_{j,3} } \right)$, following from the standard argument on
joint typicality {\cite[Theorem $15.2.1$]{info_cover}}, for each set
of error events from \eqref{equ:error:event31} to
\eqref{equ:error:event33}, we have:
\begin{align}
 &\Pr \left( {E_{j,31} } \right) \le 2^{nR_c } 2^{n\left( {\tilde R_c
+ \tilde R_p } \right)} 2^{ - n\left( {I\left( {X,X_R ,U;Y\left| V
\right.} \right) - \epsilon} \right)}, \\
 & \Pr \left( {E_{j,32} } \right) \le 2^{nR_p } \big(2^{ - n\left( {I\left( {X;Y\left| {X_R ,U,V} \right.} \right) - \epsilon} \right)}  + 2^{n\tilde R_p } 2^{ - n\left( {I\left( {X,X_R ;Y\left| {U,V} \right.} \right) - \epsilon} \right)}  \nonumber\\
 &\;\;\;\;+ 2^{n\left( {\tilde R_c  + \tilde R_p } \right)} 2^{ - n\left( {I\left( {X,X_R ,U;Y\left| V \right.} \right) - \epsilon} \right)} \big), \\
 &\Pr \left( {E_{j,33} } \right) \le 2^{n\left( {R_c  + R_p } \right)} 2^{n\left( {\tilde R_c  + \tilde R_p } \right)} 2^{ - n\left( {I\left( {X,X_R ,U;Y\left| V \right.} \right) - \epsilon} \right)}
\end{align}
for arbitrary $\epsilon > 0$ and sufficiently large $n$.

\par Note that $
I\left( {X,X_R ,U;Y\left| V \right.} \right) = I\left( {X,X_R
;Y\left| V \right.} \right)$ because $ U \leftrightarrow \left(
{X,X_R ,V} \right) \leftrightarrow Y $ forms a Markov chain. Thus,
$\Pr \left( {E_{j,3} } \right) \le \sum\limits_{k = 1}^3 {\Pr \left(
{E_{j,3k} } \right)} \to 0$ as long as
\begin{align}
 &R_p  < I\left( {X;Y\left| {X_R ,U, V} \right.} \right), \label{equ:th1:bound5}\\
 &R_p + {\tilde R}_{p}  < I\left( {X, X_R;Y\left| {U, V} \right.} \right),
 \label{equ:th1:bound6}\\
 &R_c  + R_p  + {\tilde R}_{c} + {\tilde R}_{p} < I\left( {X,X_R ;Y\left| V \right.}
 \right) \label{equ:th1:bound7}
\end{align}
for sufficiently large $n$.
\par Therefore if bounds \eqref{equ:th1:bound1},
\eqref{equ:th1:bound2}$-$\eqref{equ:th1:bound4}, and
\eqref{equ:th1:bound5}$-$\eqref{equ:th1:bound7} are satisfied,
\\ $\Pr \left( {E_j \left| {\bigcap\limits_{i = j + 1}^b {E_i^c } }
\right.} \right) \to 0$ for all $j =1,..., b-1$ and for sufficiently
large $n$.

  \end{enumerate}
  \par Collecting bounds \eqref{equ:th1:bound1},
\eqref{equ:th1:bound2}$-$\eqref{equ:th1:bound4}, and
\eqref{equ:th1:bound5}$-$\eqref{equ:th1:bound7}, along with
\eqref{equ:th1:CSR}, \eqref{equ:th1:CRS}, $R = R_c + R_p $, $\tilde
R = {\tilde R}_{c} + {\tilde R}_{p} $, applying Fourier-Motzkin
elimination{\cite[Appendix D]{YKGamalLecture}}, and exploiting the
fact that $V \leftrightarrow S \leftrightarrow Y$ forms a Markov
chain, we establish the achievable rate given by
\eqref{equ:achievable:1}$-$\eqref{equ:constraint1:th1}.
\end{itemize}

\section{Proof of Proposition \ref{proposition:achievable2}}
\label{appendix:proposition:achievable2}

\par Consider $b$ blocks of transmission. We randomly and
independently generate codebooks for each block.
\begin{itemize}
  \item {\bf Codebook Generation}:\\
  Fix a joint distribution \\
  $p\left( {s,v,u,x,x_R ,y} \right)= p\left(s\right) p\left({v\left| s, x_R,u
\right.}\right) p\left(u\right) p\left({x\left| {u } \right.}\right)
p\left({x_R\left| {u }
\right.}\right)p\left({y\left| {s,x,x_R } \right.}\right)$.\\
Define rates $R = R_c  + R_p$ with $ 0 \le R_c \le \min \left(R,
C_{SR}\right)$, and $R_v = {\tilde R}_{c} + {\tilde R}_{p}$ with $0
\le {\tilde R}_{c} \le \min \left( {R_v ,C_{RS} } \right)$.
\begin{enumerate}
  \item For each block $j$, $j \in \left[ {1:b} \right]$, generate
  $2^{n\left( {bR_c  + {\tilde R}_{c} } \right)}$ i.i.d. sequences $u_j^n $
  according to $p \left( {u_j^n } \right) = \prod\limits_{i = 1}^n {p \left( {u_{j,i} } \right)}
  $ for the given $p\left( u \right)$. Index them as $u_j^n \left( {w_c ,t_{c,j - 1} } \right)$ with $w_c
\in \left[ {1:2^{nbR_c } } \right]$
 and $t_{c,j - 1}  \in \left[ {1:2^{n{\tilde R}_{c} } } \right]$.
  \item For each block $j$, for each $u_j^n \left( {w_c ,t_{c,j - 1} }
  \right)$, generate $2^{nbR_p } $ i.i.d. sequences $x_j^n $
 according to the conditional PMF $p \left( {x_j^n \left| {u_j^n } \right.} \right) = \prod\limits_{i = 1}^n {p \left( {x_{j,i} \left| {u_{j,i} } \right.} \right)}
 $ for the given $p\left( {x\left| u \right.} \right)$. Index them as $x_j^n \left( {w_p \left| {w_c ,t_{c,j - 1} }
\right.}
 \right)$ with $w_p  \in \left[ {1:2^{nbR_p } } \right]$.
  \item For each block $j$, for each $u_j^n \left( {w_c ,t_{c,j - 1} }
  \right)$, generate $2^{n {\tilde R}_{p} } $ i.i.d. sequences $x_{R,j}^n $
 according to the conditional PMF $p \left( {x_{R,j}^n \left| {u_j^n } \right.} \right) = \prod\limits_{i = 1}^n {p \left( {x_{R,j,i} \left| {u_{j,i} } \right.} \right)}
 $ for the given $p\left( {x_R \left| u \right.} \right)$. Index them as $x_{R,j}^n \left( {t_{p,j-1} \left| {w_c ,t_{c,j -
1} } \right.}
 \right)$ with $t_{p,j-1}  \in \left[ {1:2^{n {\tilde R}_{p} } } \right]$.
 \item For each block $j$, for each $
\left( {x_{R,j}^n \left( {t_{p,j - 1} \left| {w_c ,t_{c,j - 1} }
\right.} \right),u_j^n \left( {w_c ,t_{c,j - 1} } \right)} \right)$,
generate $2^{nR_v } $ i.i.d. sequences $v_j^n $ according
 to the conditional marginal PMF
 \begin{align}
 p \left( {v_j^n \left| {x_{R,j}^n , u_j^n} \right.} \right) = \prod\limits_{i = 1}^n {p\left( {v_{j,i} \left| {x_{R,j,i},u_{j,i} } \right.} \right)}
 \nonumber
 \end{align}
for the given $p\left( {v\left| {x_R ,u} \right.} \right)$. Index
them as $v_j^n \left( {t_{c,j} ,t_{p,j} \left| {t_{c,j - 1} ,t_{p,j
- 1} ,w_c } \right.} \right)$ with $t_{c,j} \in \left[ {1:2^{n
{\tilde R}_{c} } } \right]$ and $t_{p,j} \in \left[ {1:2^{n{\tilde
R}_{p} } } \right]$.
\end{enumerate}
  \item {\bf Encoding}:
   \par The source wishes to send the same message $
w = \left( {w_{c} ,w_{p} } \right)$ to the destination over all the
blocks. At the beginning of the first block, through conferencing
link $C_{SR}$, the common message $w_{c}$ can be perfectly conveyed
to the relay as long as $bn{R_c} \le bnC_{SR}$, which implies that
  \begin{align}
   R_c  \le C_{SR}. \label{equ:th2:CSR}
  \end{align}
  Similarly, the partial compression index $t_{c,j-1} \in \left[1: 2^{n {\tilde R}_{c}}\right]$ selected at the
  relay can always be delivered to the source through the
  conferencing link $C_{RS}$ for each $j$th block as long as
  \begin{align}
  {\tilde R}_{c} \le C_{RS}. \label{equ:th2:CRS}
  \end{align}

  \par Then we have the following encoding operations:
  \begin{enumerate}
    \item $j = 1$: The source sends out $x_1^n \left( {w_p \left| {w_c ,1} \right.}
    \right)$ while the relay sends out $x_{R,1}^n \left( {1 \left| {w_c ,1} \right.}
    \right)$.
    \item $j = \left[ {2:b} \right]$: At the end of block $\left(j-1\right)$, the relay learns the entire state sequence, i.e.,
    $s_{j - 1}^n$, and looks for a compression codeword $v_{j-1}^n$
associated with index $\left( {t_{c,j - 1} ,t_{p,j - 1} } \right)$
such that
\begin{align}
 \left( \begin{array}{l}
 s_{j - 1}^n ,v_{j - 1}^n \left( {t_{c,j - 1} ,t_{p,j - 1} \left| {t_{c,j - 2} ,t_{p,j - 2} ,w_c } \right.} \right), \\
 x_{R,j - 1}^n \left( {t_{p,j - 2} \left| {t_{c,j - 2} ,w_c } \right.} \right),u_{j - 1}^n \left( {w_c ,t_{c,j - 2} } \right) \\
 \end{array} \right) \in A_\epsilon^n. \nonumber
 \end{align}
 If more than one codewords are found, choose the first one in the
list. If there is no such a codeword, choose an arbitrary one at
random from the compression codebook. Then codeword $ x_j^n \left(
{w_{p} \left| {w_{c} ,t_{c,j-1}} \right.} \right)$ is sent out by
the source and codeword $x_{R,j}^n \left( {t_{p,j-1} \left| {w_c
,t_{c,j-1}} \right.} \right)$ is sent out by the relay.
  \end{enumerate}
  \item {\bf Decoding}:
  \par At the end of $b$ blocks of transmission, the destination
  performs joint decoding over all blocks by looking for an unique
  message $\hat w = \left( {\hat w_{c} ,\hat
w_{p} } \right)$ with $\hat w_c  \in \left[ {1:2^{nbR_c } } \right]$
and $\hat w_p  \in \left[ {1:2^{nbR_p } } \right]$ such that:
\begin{align}
\left( \begin{array}{l}
 x_j^n \left( {\hat w_p \left| {\hat w_c } \right.,t_{c,j - 1} } \right),x_{R,j}^n \left( {t_{p,j - 1} \left| {\hat w_c ,t_{j - 1} } \right.} \right), \\
 v_j^n \left( {t_{c,j} ,t_{p,j} \left| {t_{c,j - 1} ,t_{p,j - 1} ,\hat w_c } \right.} \right),u_j^n \left( {\hat w_c ,t_{c,j - 1} } \right),y_j^n  \\
 \end{array} \right) \in A_\epsilon^n
\end{align}
for all $j = 1,...,b$ and some ${\bf{t}}^b  \buildrel \Delta \over =
\left( {{\bf{t}}_1 ,{\bf{t}}_2 ,...,{\bf{t}}_b } \right) = \left(
{t_{c,1} ,t_{p,1} ,t_{c,2} ,t_{p,2} ,...,t_{c,b} ,t_{p,b} }
\right)$.

  \item {\bf Analysis of Probability of Error}:
   \par To bound the probability of error $\Pr(E)$, WLOG, assume $\left(w_c, w_p\right) = \left(1, 1 \right) $ are sent for all blocks.
   Also denote the indices selected by the relay
   for each block by $\left(T_{c,j-1}, T_{p,j-1}\right)$.
  \par Define the following encoding error events:
  \begin{align}
E_0  = \bigcup\limits_{j = 1}^b {\left\{ \begin{array}{l}
 \left( {v_j^n \left( {t_{c,j} ,t_{p,j} \left| {T_{c,j - 1} ,T_{p,j - 1} ,1} \right.} \right),s_j^n ,x_{R,j}^n \left( {T_{p,j - 1} \left| {T_{c,j - 1} ,1} \right.} \right),u_j^n \left( {1,T_{c,j - 1} } \right)} \right) \notin A_{\epsilon}^n , \\
 \forall \:t_{c,j}  \in \left[ {1:2^{n {\tilde R}_{c} } } \right],\forall \:t_{p,j}  \in \left[ {1:2^{n {\tilde R}_{p} } } \right] \\
 \end{array} \right\}}
  \end{align}
  \par Define the following decoding events:
  \begin{align}
 E_{\left( {w_c ,w_p } \right)} = \left\{ {\begin{array}{*{20}c}
   {\bigcap\limits_{j = 1}^b {A_j \left( {w_c ,w_p ,t_{c,j} ,t_{p,j}
,t_{c,j - 1} ,t_{p,j - 1} } \right)}} \\
   {{\rm{for}}\:{\rm{some}}\:\:{\bf{t}}^b  = \left( {t_{c,1} ,t_{p,1} ,t_{c,2} ,t_{p,2} ,...,t_{c,b} ,t_{p,b} } \right).}  \\
\end{array}} \right\},
  \end{align}
  where each $A_j \left( {w_c ,w_p ,t_{c,j} ,t_{p,j} ,t_{c,j - 1} ,t_{p,j - 1} }
  \right)$ is defined by:
   \begin{align}
   &A_j \left( {w_c ,w_p ,t_{c,j} ,t_{p,j} ,t_{c,j - 1} ,t_{p,j - 1} } \right) \nonumber \\
   &\buildrel \Delta \over = \left\{ {\left( {\begin{array}{*{20}c}
   {x_j^n \left( {w_p \left| {w_c ,t_{c,j - 1} } \right.} \right),x_{R,j}^n \left( {t_{p,j - 1} \left| {w_c ,t_{c,j - 1} } \right.} \right),}  \\
   {v_j^n \left( {t_{c,j} ,t_{p,j} \left| {t_{c,j - 1} ,t_{p,j - 1} ,w_c } \right.} \right),u_j^n \left( {w_c ,t_{c,j - 1} } \right),y_j^n }  \\
\end{array}} \right) \in A_{\epsilon}^n } \right\}.
   \end{align}

  \par Hence by the union bound,
  \begin{align}
 \Pr \left( E \right) &\le \Pr \left( {E_0 } \right) + \Pr \left( {E_{\left( {1,1} \right)}^c  \cap E_0^c } \right) + \Pr \left( {\bigcup\limits_{\left( {w_c ,w_p } \right) \ne \left( {1,1} \right)} {E_{\left( {w_c ,w_p } \right)} } } \right) \\
  &\le \Pr \left( {E_0 } \right) + \Pr \left( {E_{\left( {1,1} \right)}^c  \cap E_0^c } \right) + \sum\limits_{w_c  \ne 1,w_p  \ne 1} {\Pr \left( {E_{\left( {w_c ,w_p } \right)} } \right)} \nonumber \\
  &\;\;\;\;\;\;\;\;\;\;+ \sum\limits_{w_c  \ne 1} {\Pr \left( {E_{\left( {w_c, 1 } \right)} } \right)}  + \sum\limits_{w_p  \ne 1} {\Pr \left( {E_{\left( {1, w_p} \right)} }
  \right)}.
  \end{align}

  \begin{enumerate}
    \item By the covering lemma in
    \cite{YKGamalLecture}, $\Pr \left( {E_0 } \right) \to 0$ as long as
    \begin{align}
R_v  > I\left( {V;S\left| {X_R, U } \right.} \right)
\label{equ:th2:bound1}
    \end{align}
    for sufficiently large $n$.
    \item By the conditional joint typicality lemma in \cite{YKGamalLecture}, $
\Pr \left( {E_{\left( {1,1} \right)}^c \left| {E_0^c } \right.}
\right) \to 0 $ for sufficiently large $n$.
    \item For $\sum\limits_{w_c  \ne 1,w_p  \ne 1} {\Pr \left( {E_{\left( {w_c ,w_p } \right)} }
    \right)}$, we have
        \begin{align}
  &\sum\limits_{w_c  \ne 1,w_p  \ne 1} {\Pr \left( {E_{\left( {w_c ,w_p } \right)} } \right)}  \nonumber \\
  &= \sum\limits_{w_c  \ne 1,w_p  \ne 1} {\Pr \left( {\bigcup\limits_{{\bf{t}}^b } {\bigcap\limits_{j = 1}^b {A_j \left( {w_c ,w_p ,t_{c,j}, t_{p,j} ,t_{c,j - 1},t_{p,j - 1} } \right)} } } \right)}  \label{equ:th2:erp:1}\\
  &\le \sum\limits_{w_c  \ne 1,w_p  \ne 1} {\sum\limits_{{\bf{t}}^b } {\Pr \left( {\bigcap\limits_{j = 1}^b {A_j \left( {w_c ,w_p ,t_{c,j}, t_{p,j} ,t_{c,j - 1},t_{p,j - 1} } \right)} } \right)} }  \label{equ:th2:erp:2}\\
  &= \sum\limits_{w_c  \ne 1,w_p  \ne 1} {\sum\limits_{{\bf{t}}^b } {\prod\limits_{j = 1}^b {\Pr \left( {A_j \left( {w_c ,w_p ,t_{c,j}, t_{p,j} ,t_{c,j - 1},t_{p,j - 1} } \right)} \right)} } }  \label{equ:th2:erp:3}\\
  &\le \sum\limits_{w_c  \ne 1,w_p  \ne 1} {\sum\limits_{{\bf{t}}^b }
{\prod\limits_{j = 2}^b {\Pr \left( {A_j \left( {w_c ,w_p ,t_{c,j}
,t_{p,j} ,t_{c,j - 1} ,t_{p,j - 1} } \right)} \right)} } }
\label{equ:th2:erp:4}\\
  &\le 2^{nb\left( {R_c  + R_p } \right)}2^{n{b}\left({\tilde R}_{c} + {\tilde R}_{p}\right)} 2^{ - n\left( {b - 1} \right)\left( {I\left( {X,X_R ,V, U;Y} \right) - \epsilon }
  \right)} \label{equ:th2:erp:5}
    \end{align}
where \eqref{equ:th2:erp:2} holds by the union bound;
\eqref{equ:th2:erp:3} holds due to the independence of codebook for
each block and the memoryless property of the channel;
\eqref{equ:th2:erp:4} follows from $0 \le \Pr\left( {A_j } \right)
\le 1$; and \eqref{equ:th2:erp:5} follows from the fact that
\begin{align}
\Pr \left( {A_j \left( {w_c ,w_p ,t_{c,j}, t_{p,j} ,t_{c,j -
1},t_{p,j - 1} } \right)} \right) \le 2^{ - n\left( {I\left( {X,X_R
,V,U;Y} \right) - \epsilon}\right)} \nonumber
\end{align}
for arbitrary
$\epsilon
> 0$ and sufficiently large $n$ when $ {w_c \ne 1}$ and $w_p \ne 1$,
by the standard argument on joint typicality {\cite[Theorem
$15.2.1$]{info_cover}}.
\par Thus
$\sum\limits_{w_c  \ne 1,w_p  \ne 1} {\Pr \left( {E_{\left( {w_c
,w_p } \right)} }
    \right)} \to 0$ as long as
    \begin{align}
R_c  + R_p  < \frac{{b - 1}}{b}\left( {I\left( {X,X_R,V,U;Y} \right)
- \epsilon} \right) - \left({\tilde R}_{c} + {\tilde R}_{p}\right)
    \end{align}
    for arbitrary $\epsilon >0$ and sufficiently large $n$.
\par Setting $b \to \infty$, we have
    \begin{align}
R_c  + R_p  < I\left( {X,X_R,V,U;Y} \right) - \left({\tilde R}_{c} +
{\tilde R}_{p}\right). \label{equ:th2:bound2}
    \end{align}

\item For $ \sum\limits_{w_c  \ne 1} {\Pr \left( {E_{\left( {w_c,1 } \right)} }
  \right)}$, following similar arguments from \eqref{equ:th2:erp:1}
  to \eqref{equ:th2:erp:4}, we have
  \begin{align}
  &\sum\limits_{w_c  \ne 1} {\Pr \left( {E_{\left( {w_c,1 } \right)} } \right)}  \nonumber \\
  &\le \sum\limits_{w_c  \ne 1} {\sum\limits_{{\bf{t}}^b } {\prod\limits_{j
= 2}^b {\Pr \left( {A_j \left( {w_c ,1, t_{c,j} ,t_{p,j} ,t_{c,j -
1} ,t_{p,j - 1} } \right)} \right)} } } \label{equ:th2:erp1:1} \\
  &\le 2^{nR_c } 2^{n{b} \left({\tilde R}_{c} + {\tilde R}_{p}\right) } 2^{ - n\left( {b - 1} \right)\left( {I\left( {X,X_R,V,U;Y} \right) - \epsilon}
  \right)} \label{equ:th2:erp1:4}
  \end{align}
  where \eqref{equ:th2:erp1:4} follows from the fact that
\begin{align}
\Pr \left( {A_j \left( {w_c ,1, t_{c,j}, t_{p,j} ,t_{c,j - 1},
t_{p,j - 1} } \right)} \right) \le  2^{ - n\left( {I\left( {X,X_R
,V,U;Y} \right) - \epsilon} \right)} \nonumber
\end{align}
for arbitrary $\epsilon > 0$
 and sufficiently large $n$ when $w_c \ne 1$, by the
standard argument on joint typicality {\cite[Theorem
$15.2.1$]{info_cover}}.

\par Thus $\sum\limits_{w_c  \ne 1} {\Pr \left( {E_{\left( {w_c,1 } \right)} }
  \right)} \to 0$ as long as
    \begin{align}
R_c < \frac{{b - 1}}{b}\left( {I\left( {X,X_R,V,U;Y} \right) -
\epsilon} \right) - \left({\tilde R}_{c} + {\tilde R}_{p}\right)
    \end{align}
    for arbitrary $\epsilon >0$ and sufficiently large $n$.
\par Setting $b \to \infty$, we have
    \begin{align}
R_c < I\left( {X,X_R,V,U;Y} \right) - \left({\tilde R}_{c} + {\tilde
R}_{p}\right).
    \end{align}
 \par But notice that \eqref{equ:th2:bound2} implies this bound,
 hence it is redundant.

 \item For $\sum\limits_{w_p  \ne 1} {\Pr \left( {E_{\left( {1,w_p} \right)} }
 \right)}$, again following similar arguments from \eqref{equ:th2:erp:1}
  to \eqref{equ:th2:erp:4}, we have
  \begin{align}
  &\sum\limits_{w_p  \ne 1} {\Pr \left( {E_{\left( {1,w_p } \right)} } \right)}  \nonumber \\
  &\le \sum\limits_{w_p  \ne 1} {\sum\limits_{{\bf t}_b } {\sum\limits_{{\bf{t}}^{b - 1} } {\prod\limits_{j = 2}^b {\Pr \left( {A_j \left( {1,w_p ,t_{c,j} ,t_{p,j}, t_{c,j - 1},t_{p,j - 1} } \right)} \right)} } } }
  \label{equ:th2:erp2:1}
  \end{align}

  \par By the standard argument
on joint typicality {\cite[Theorem $15.2.1$]{info_cover}} for
enumerations over all $ \left(t_{c,j} ,t_{p,j}, t_{c,j - 1} ,t_{p,j
- 1}\right)$ given any fixed $ \left( {w_c  = 1, w_p  \ne 1}
\right)$, we have
\begin{align}
 &\Pr \left( {A_j \left( {1, w_p,t_{c,j},t_{p,j},t_{c,j - 1},t_{p,j - 1} } \right)} \right) \\
  &\le \left\{ {\begin{array}{*{20}c}
   {2^{ - n\left( {I\left( {X;Y\left| {X_R ,V, U} \right.} \right) - \epsilon} \right)} ,
 \;{\rm{if}}\;{t_{c,j - 1}  = 1, t_{p,j - 1}  = 1\;{\rm{and\;for\;any}}\;{t_{c,j}, t_{p,j}} }}  \\
   {2^{ - n\left( {I\left( {X,X_R ,V;Y\left| { U} \right.} \right) - \epsilon} \right)} , \;{\rm{if}}\;{t_{c,j - 1} = 1, t_{p,j - 1}  \ne 1 \;{\rm{and\;for\;any}}\;{t_{c,j}, t_{p,j}} } } \\
   {2^{ - n\left( {I\left( {X,X_R,V,U;Y} \right) - \epsilon} \right)},{\;{\rm{if }}\; t_{c,j - 1}\ne 1\;{\rm{and\;for\;any}}\;t_{p,j - 1},t_{c,j},t_{p,j} }\;\;\;\;\;\;\;\;\; } \\
\end{array}} \right. \\
  &\buildrel \Delta \over = {Q_j \left( {1, w_p,t_{c,j - 1} ,t_{p,j - 1} } \right)}
\end{align}
where the upper bound is dependent on $\left(t_{c,j-1},
t_{p,j-1}\right)$ only, for arbitrary $\epsilon
> 0$ and sufficiently large $n$. Then we have
\begin{align}
 &\sum\limits_{{\bf{t}}_{j - 1} } {Q_j \left( {1, w_p,t_{c,j - 1} ,t_{p,j - 1} } \right)}  \\
 &\le 2^{ - n\left( {I\left( {X;Y\left| {X_R ,V,U} \right.} \right) - \epsilon} \right)}  + 2^{n {\tilde R}_p } 2^{ - n\left( {I\left( {X,X_R ,V;Y\left| U \right.} \right) - \epsilon} \right)}  + 2^{n\left( {{\tilde R}_c  + {\tilde R}_p } \right)} 2^{ - n\left( {I\left( {X,X_R,V,U;Y} \right) - \epsilon}
 \right)}\\
  &\le 3 \times 2^{ - n\left( {\min \left( {{\bf I}_1 ,{\bf I}_2 ,{\bf I}_3 } \right) - \epsilon} \right)}
\end{align}
with
  \begin{align}
 &{\bf I}_1  = I\left( {X;Y\left| {X_R ,V, U} \right.} \right), \\
 &{\bf I}_2  = I\left( {X,X_R,V;Y\left| { U} \right.} \right) - {\tilde R}_{p}, \\
 &{\bf I}_3  = I\left( {X,X_R,V,U;Y} \right) - \left( {{\tilde R}_{c}  + {\tilde R}_{p} }
 \right),
  \end{align}
  for arbitrary $\epsilon > 0$ and sufficiently large $n$.
\par Hence,
  \begin{align}
  &\sum\limits_{w_p  \ne 1} {\Pr \left( {E_{\left( {1, w_p} \right)} }
 \right)} \nonumber \\
    &\le \sum\limits_{w_p  \ne 1} {\sum\limits_{{\bf t}_b } {\sum\limits_{{\bf{t}}^{b - 1} } {\prod\limits_{j = 2}^b {Q_j \left( {1, w_p
,t_{c,j - 1} ,t_{p,j - 1} } \right)} } } }  \label{equ:th2:erp2:2} \\
  &= \sum\limits_{w_p  \ne 1} {\sum\limits_{{\bf t}_b } {\prod\limits_{j = 2}^b {\sum\limits_{{\bf t}_{j - 1} } {Q_j \left( {1, w_p
,t_{c,j - 1} ,t_{p,j - 1} } \right)} } }
 } \label{equ:th2:erp2:3} \\
  &\le 2^{nbR_p } 2^{n\left({\tilde R}_{c} + {\tilde R}_{p}\right) } 3^{\left( {b - 1} \right)} 2^{
 - n\left( {b - 1} \right)\left( {\min \left( {\bf I}_1, {\bf I}_2, {\bf I}_3 \right) - \epsilon } \right)} \label{equ:th2:erp2:4}
  \end{align}
where \eqref{equ:th2:erp2:3} holds because ${Q_j \left( {1,
w_p,t_{c,j - 1} ,t_{p,j - 1} } \right)}$ is dependent on
$\left(t_{c,j-1}, t_{p,j-1}\right)$ only when $w_p \ne 1$.

\par Thus $\sum\limits_{w_p  \ne 1} {\Pr \left( {E_{\left( {1, w_p } \right)} }
  \right)} \to 0$ as long as
    \begin{align}
R_p  < \frac{{b - 1}}{b}\left( {\min \left( {\bf I}_1, {\bf I}_2,
{\bf I}_3 \right) - \epsilon } \right) - \frac{1}{b}\left({\tilde
R}_{c}+ {\tilde R}_{p}\right) - \frac{{\left(b - 1\right)}{\log _2
3}}{{nb }}
    \end{align}
    for arbitrary $\epsilon >0$ and sufficiently large $n$.
\par Setting $b \to \infty$ and $n \to \infty$, we have
    \begin{align}
 R_p &< {\bf I}_1 = I\left( {X;Y\left| {X_R ,V,U} \right.} \right),
     \label{equ:th2:bound3}\\
 R_p &< {\bf I}_2 = I\left( {X,X_R,V;Y\left| {U} \right.} \right) - {\tilde R}_{p},  \label{equ:th2:bound4}\\
 R_p &< {\bf I}_3 = I\left( {X,X_R,V,U;Y} \right) - \left( {{\tilde R}_{c}  + {\tilde R}_{p} }
 \right).\label{equ:th2:boundredundant}
    \end{align}
\par Again notice that \eqref{equ:th2:bound2} implies \eqref{equ:th2:boundredundant},
 hence \eqref{equ:th2:boundredundant} is redundant.
   \end{enumerate}
   Collecting all the necessary constraints \eqref{equ:th2:CSR},
   \eqref{equ:th2:CRS},
\eqref{equ:th2:bound1}, \eqref{equ:th2:bound2},
\eqref{equ:th2:bound3} and \eqref{equ:th2:bound4}, combining with $R
= R_c  + R_p$, $R_v  = {\tilde R}_{c}  + {\tilde R}_{p}$, and
applying Fourier-Motzkin elimination{\cite[Appendix
D]{YKGamalLecture}}, we finally establish the achievable rate given
by \eqref{equ:achievable:2}$-$\eqref{equ:inputd:2}.
\end{itemize}

\bibliographystyle{IEEEtran}
\bibliography{RelayWithState}

\begin{thebibliography}{10}
\providecommand{\url}[1]{#1}
\csname url@samestyle\endcsname
\providecommand{\newblock}{\relax}
\providecommand{\bibinfo}[2]{#2}
\providecommand{\BIBentrySTDinterwordspacing}{\spaceskip=0pt\relax}
\providecommand{\BIBentryALTinterwordstretchfactor}{4}
\providecommand{\BIBentryALTinterwordspacing}{\spaceskip=\fontdimen2\font plus
\BIBentryALTinterwordstretchfactor\fontdimen3\font minus
  \fontdimen4\font\relax}
\providecommand{\BIBforeignlanguage}[2]{{%
\expandafter\ifx\csname l@#1\endcsname\relax
\typeout{** WARNING: IEEEtran.bst: No hyphenation pattern has been}%
\typeout{** loaded for the language `#1'. Using the pattern for}%
\typeout{** the default language instead.}%
\else
\language=\csname l@#1\endcsname
\fi
#2}}
\providecommand{\BIBdecl}{\relax}
\BIBdecl

\bibitem{shannon1958channels}
C.~E. Shannon, ``{Channels with side information at the transmitter},''
  \emph{IBM Journal of Research and Development}, vol.~2, no.~4, pp. 289--293,
  October 1958.

\bibitem{gel1980coding}
S.~I. Gel'Fand and M.~S. Pinsker, ``{Coding for channel with random
  parameters},'' \emph{Problems of Control and Information Theory}, vol.~9,
  no.~1, pp. 19--31, 1980.

\bibitem{keshet2007channel}
G.~Keshet, Y.~Steinberg, and N.~Merhav, ``{Channel coding in the presence of
  side information},'' \emph{Foundations and Trends in Communications and
  Information Theory}, vol.~4, no.~6, pp. 445--586, 2007.

\bibitem{LS_IZS2010}
A.~Lapidoth and Y.~Steinberg, ``{The multiple access channel with causal and
  strictly causal side information at the encoders},'' in \emph{Proceedings of
  International Zurich Seminar on Communications}, March 2010.

\bibitem{LS_ISIT2010}
------, ``The multiple access channel with two independent states each known
  causally to one encoder,'' in \emph{Proceedings of IEEE International
  Symposium on Information Theory}, June 2010.

\bibitem{DPC_1983}
M.~Costa, ``Writing on dirty paper,'' \emph{IEEE Transactions on Information
  Theory}, vol.~29, no.~3, pp. 439--441, May 1983.

\bibitem{sigurjonsson2005multiple}
S.~Sigurjonsson and Y.~H. Kim, ``{On multiple user channels with state
  information at the transmitters},'' in \emph{Proceedings of IEEE
  International Symposium on Information Theory}, September 2005.

\bibitem{jafar2006capacity}
S.~Jafar, ``{Capacity with causal and noncausal side information: A unified
  view},'' \emph{IEEE Transactions on Information Theory}, vol.~52, no.~12, pp.
  5468--5474, December 2006.

\bibitem{philosof2007lattice}
T.~Philosof, A.~Khisti, U.~Erez, and R.~Zamir, ``{Lattice strategies for the
  dirty multiple access channel},'' in \emph{Proceedings of IEEE International
  Symposium on Information Theory}, June 2007.

\bibitem{somekh2008cooperative}
A.~Somekh-Baruch, S.~Shamai~(Shitz), and S.~Verdu, ``{Cooperative
  multiple-access encoding with states available at one transmitter},''
  \emph{IEEE Transactions on Information Theory}, vol.~54, no.~10, pp.
  4448--4469, October 2008.

\bibitem{kotagiri2008multiaccess}
S.~P. Kotagiri and J.~N. Laneman, ``{Multiaccess channels with state known to
  some encoders and independent messages},'' \emph{EURASIP Journal on Wireless
  Communications and Networking}, vol. 2008, pp. 1--14, January 2008.

\bibitem{Permuter2010}
H.~Permuter, S.~Shamai~(Shitz), and A.~Somekh-Baruch, ``Message and state
  cooperation in multiple access channels,'' 2010, available online at
  http://arxiv.org/abs/1006.2022.

\bibitem{mirmohseni2009compress}
M.~Mirmohseni, B.~Akhbari, and M.~R. Aref, ``{Compress-and-forward strategy for
  the relay channel with causal state information},'' in \emph{Proceedings of
  IEEE Information Theory Workshop}, October 2009.

\bibitem{akhbari2010state}
B.~Akhbari, M.~Mirmohseni, and M.~R. Aref, ``{State-dependent relay channel
  with private messages with partial causal and non-causal channel state
  information},'' in \emph{Proceedings of IEEE International Symposium on
  Information Theory}, June 2010.

\bibitem{zaidi2010cooperative}
A.~Zaidi, S.~P. Kotagiri, J.~N. Laneman, and L.~Vandendorpe, ``{Cooperative
  relaying with state available noncausally at the relay},'' \emph{IEEE
  Transactions on Information Theory}, vol.~56, no.~5, pp. 2272--2298, May
  2010.

\bibitem{zaidi2010bounds}
A.~Zaidi, S.~Shamai~(Shitz), P.~Piantanida, and L.~Vandendorpe, ``{Bounds on
  the capacity of the relay channel with noncausal state information at
  source},'' in \emph{Proceedings of IEEE International Symposium on
  Information Theory}, June 2010.

\bibitem{cover1979capacity}
T.~M. Cover and A.~El~Gamal, ``{Capacity theorems for the relay channel},''
  \emph{IEEE Transactions on Information Theory}, vol.~25, no.~5, pp. 572--584,
  September 1979.

\bibitem{shannon1956zero}
C.~E. Shannon, ``{The zero error capacity of a noisy channel},'' \emph{IRE
  Transactions on Information Theory}, vol.~2, no.~3, pp. 8--19, September
  1956.

\bibitem{lisimeoneyener2010}
M.~Li, O.~Simeone, and A.~Yener, ``{Multiple access channels with states
  causally known at transmitters},'' November 2010, submitted to {\it IEEE
  Transactions on Information Theory}, available online at
  http://arxiv.org/abs/1011.6639.

\bibitem{lim2010noisy}
S.~H. Lim, Y.~H. Kim, A.~El~Gamal, and S.~Y. Chung, ``{Noisy network coding},''
  2010, available online at http://arxiv.org/abs/1002.3188v2.

\bibitem{elgamal2005capacity}
A.~El~Gamal and S.~Zahedi, ``{Capacity of a class of relay channels with
  orthogonal components},'' \emph{IEEE Transactions on Information Theory},
  vol.~51, no.~5, pp. 1815--1817, May 2005.

\bibitem{willems1983discrete}
F.~Willems, ``{The discrete memoryless multiple access channel with partially
  cooperating encoders},'' \emph{IEEE Transactions on Information Theory},
  vol.~29, no.~3, pp. 441--445, May 1983.

\bibitem{dabora2006broadcast}
R.~Dabora and S.~D. Servetto, ``{Broadcast channels with cooperating
  decoders},'' \emph{IEEE Transactions on Information Theory}, vol.~52, no.~12,
  pp. 5438--5454, December 2006.

\bibitem{bross2008gaussian}
S.~I. Bross, A.~Lapidoth, and M.~A. Wigger, ``{The Gaussian MAC with
  conferencing encoders},'' in \emph{Proceedings of IEEE International
  Symposium on Information Theory}, July 2008.

\bibitem{simeone2008three}
O.~Simeone, O.~Somekh, G.~Kramer, H.~V. Poor, and S.~Shamai~(Shitz),
  ``{Three-user Gaussian multiple access channel with partially cooperating
  encoders},'' in \emph{Proceedings of the 42nd Asilomar Conference on Signals,
  Systems and Computers}, October 2008.

\bibitem{gesbertmulti2010}
D.~Gesbert, S.~Hanly, H.~Huang, S.~Shitz, O.~Simeone, and W.~Yu, ``{Multi-cell
  MIMO cooperative networks: A new look at interference},'' \emph{IEEE Journal
  on Selected Areas in Communications}, vol.~28, no.~9, pp. 1380--1408,
  December 2010.

\bibitem{chandrasekhar2008femtocell}
V.~Chandrasekhar, J.~Andrews, and A.~Gatherer, ``{Femtocell networks: a
  survey},'' \emph{IEEE Communications Magazine}, vol.~46, no.~9, pp. 59--67,
  September 2008.

\bibitem{willems1982informationtheoretical}
F.~Willems, ``{Information theoretical results for the discrete memoryless
  multiple access channel},'' \emph{PhD Thesis, Katholieke Univ. Leuven,
  Leuven, Belgium}, 1982.

\bibitem{wynerziv1976}
A.~Wyner and J.~Ziv, ``{The rate-distortion function for source coding with
  side information at the decoder},'' \emph{IEEE Transactions on Information
  Theory}, vol.~22, no.~1, pp. 1--10, January 1976.

\bibitem{info_cover}
T.~M. Cover and J.~A. Thomas, \emph{Elements of Information Theory}.\hskip 1em
  plus 0.5em minus 0.4em\relax Wiley-Interscience, July 2006.

\bibitem{YKGamalLecture}
A.~El~Gamal and Y.~H. Kim, ``{Lecture notes on network information theory},''
  2010, available online at http://arxiv.org/abs/1001.3404.

\bibitem{aleksic2009capacity}
M.~Aleksic, P.~Razaghi, and W.~Yu, ``{Capacity of a class of modulo-sum relay
  channels},'' \emph{IEEE Transactions on Information Theory}, vol.~55, no.~3,
  pp. 921--930, March 2009.

\bibitem{SKfeedback}
J.~Schalkwijk and T.~Kailath, ``{A coding scheme for additive noise channels
  with feedback--I: No bandwidth constraint},'' \emph{IEEE Transactions on
  Information Theory}, vol.~12, no.~2, pp. 172 -- 182, April 1966.

\end{thebibliography}

\end{document}